\documentclass{aa}  

\usepackage{graphicx}
\usepackage{txfonts}
\usepackage{natbib}
\bibpunct{(}{)}{;}{a}{}{,} 
\begin{document}

   \title{Orbital analysis of the Pluto-Charon moon system's mutual interactions and forced frequencies}
   
   \titlerunning{Frequencies in Pluto-Charon moon system}


   \author{Dionysios Gakis \thanks{dgakis@upnet.gr}
          \and
          Konstantinos N. Gourgouliatos \thanks{kngourg@upatras.gr}
          }
          
 \institute{Department of Physics, University of Patras, Patras, Rio, 26504, Greece\\
             }
             
   \date{Received 9 August 2022; accepted 4 January 2023}

 
  \abstract
   {The orbits of the four small moons in the Pluto-Charon system, Styx, Nix, Kerberos, and Hydra, are 
circumbinary, as Pluto and Charon form a binary dwarf planet. Consequently, the orbit of each moon is characterized by a number of frequencies, arising from the central binary and the mutual gravitational interactions.}
   {In this work, we identify the most prominent of these forced frequencies using fast Fourier transforms.}
   {Two methods were implemented, a semi-analytic and a numerical one, and comparisons are made.}
   {The results indicate that as a first approximation, moon orbits may well be modeled as the superposition of a series of inevitable oscillations induced by Pluto and Charon, deviating from circular orbits, even if the eccentricity is set to zero. Moreover, the mutual gravitational effects are significant in their long-term evolution, especially for the lighter moons Styx and Kerberos, activating 
modes that dominate the low-frequency region of the power spectrum. This becomes evident through the comparison of simulations where only one moon is included along with the binary dwarf planet and simulations of the entire six-body system. These modes become noticeable over long integration times and may affect the orbits of the lighter moons of the system.}
   {}


   \maketitle
%

\section{Introduction}
\label{sec:1} 

Pluto's moon system is a dynamical treasure. As the mass ratio of the dwarf planet Pluto to its largest moon Charon is 8:1 \citep{Stern:2015}, they are in fact a binary dwarf planet. Along with the central binary, with at present four known moons orbiting the system's center of mass, namely Styx, Nix, Kerberos, and Hydra, this structure is valuable for studying circumbinary orbits in depth. Thus, studying the motions of these small moons is of particular interest, as the potential arising from Pluto and Charon forces them into orbits that deviate significantly from the standard elliptical ones. 

Circumbinary orbits differ greatly from those described by Keplerian orbital elements. \cite{Lee:2006} developed a theoretical solution to modeling orbits around a zero-eccentricity binary system. Their theory, which holds for point masses on circumbinary coplanar orbits, yields that a circumbinary orbit is the superposition of a circular orbit around the center of mass, an epicyclic motion caused by the binary and a vertical component. \cite{Leung:2013} generalized this theory to include eccentric orbits of the central binary as well.			

Another study, by \cite{Bromley-Kenyon:2020}, revisited the above theory and provided quantitative tools to apply it in practice. A “most circular” circumbinary orbit is defined, corresponding to a circular orbit around a single mass. Deviations from the most circular orbit are quantified using the free eccentricity ($e_{free}$). They tested outcomes for the eccentricity damping of tracer particles in the Pluto-Charon system, along with other extrasolar planetary systems, achieving their objective satisfactorily. Nevertheless, it is acknowledged that more precise techniques are required to analyze the actual moon orbits instead of test particles. In a follow-up study, \cite{2022AJ....163..238K} further examined and set improved constraints on the dynamical behavior and masses of the smallest moons by performing an array of simulations.

\cite{Woo:2020} used fast Fourier transforms (FFTs) on numerical simulations of the dynamical system to estimate the exact values of the amplitudes and frequencies that outline the peculiar orbits of the moons of Pluto and Charon. Although they confirmed the accuracy of this method, they found significant deviations from the expected orbit of Styx. By adopting the Hamiltonian approach of \cite{Lithwick:2008}, they propose that at least a part of these deviations can be explained by the 3:1 mean motion resonance.

Some other theoretical solutions for circumbinary orbits exist as well. For instance, \cite{Georgakarakos:2015} used perturbations of the Runge–Lenz vector to study the short-term of the evolution of low-eccentricity orbits in a hierarchical triple system. Another example appears in the work of \cite{Sutherland:2019}, who proposed using empirical geometric orbital elements to search for active resonances in orbits around a binary system. These studies do result in compatible solutions to those based on the \cite{Lee:2006} theory, so we do not discuss them further.

In a previous paper \citep{Gakis:2022}, we examined the moon motions within the dynamical system of Pluto and Charon. Despite attempting to define orbits that are as close as possible to circular, moons, nonetheless appeared to deviate from such orbits, and the barycentric distance varied significantly. Our conclusion was that the time-dependent, non-axisymmetric potential of Pluto and Charon induces irregular patterns in the orbits. We also inspected that major discrepancies concerning Keplerian orbital elements between different studies so far do not primarily reflect limitations and inaccuracies in measurements, observations and calculations, but are in fact a result of the underlying specifications of the actual system.

This work is the second part of our analysis concerning the circumbinary orbits within the gravitational system of Pluto and Charon. Here, our goal is to give a more quantitative view on the dynamical specifications of the system, focusing on the effect of mutual interactions between the moons, in addition to the impact of the central binary that we have already studied. To this end, we utilize both semi-analytic and numerical approaches, and infer the most prominent amplitudes and frequencies. Specifically, we apply FFTs to identify the exact frequencies of the many oscillations that moons perform. This way we quantify the impact of the mutual interactions between the moons that force additional frequencies in the orbits.
 
The structure of this paper is the following. In Section \ref{sec:2}, we describe our calculations, which are both semi-analytic (Section \ref{sec:2.1}) and numerical (Section \ref{sec:2.2}). Section \ref{sec:3} contains our main results. Our final conclusions are summarized in the last section (Section \ref{sec:4}).

\section{Calculations}
\label{sec:2} 


The orbital elements of the dynamical system, used in this work, are presented in Table~\ref{tab:1}. There are notable differences between the data set of \cite{Showalter:2015}, and another prominent study, by \cite{Brozovic:2015}. These deviations are most probably explained by the intrinsic behavior of the system, as illustrated by \cite{Gakis:2022}. In particular, observations at different time periods inevitably produced distinct outcomes because of the variations in the relative positions of the bodies.

\subsection{Semi-analytic approach}
\label{sec:2.1} 

A short description of the model for circumbinary orbits, introduced by \cite{Lee:2006} and reconsidered by \cite{Bromley-Kenyon:2020}, on which our semi-analytic approach is principally based, is given below. Adopting cylindrical coordinates and assuming that the barycenter lies on the origin $O\,(0, 0, 0)$, the potential caused by the central binary system at a point $P\,(R, \phi, z=0)$ may be approximated by a cosine series: 
\begin{equation}\label{1} 
\Phi\,(R, \phi, z=0)=\sum_{k=0}^\infty \Phi_{k} \cos k\left(\phi-n_{PC}t\right)\,.
\end{equation}
The coefficients $\Phi_{k}$ are related to the binary properties (masses $M_P$ and $M_C$ and separation $a_{PC}$) and the orbital radius of the moon, $R_S$. The orbital frequency of Pluto and Charon is given by
\begin{equation}\label{7} n_{PC}=\sqrt{\frac{GM}{a_{PC}^3}}\,.
\end{equation}
Solving the equations of motion, the solution yields the following expressions for the position of a point-mass body initially located at $P$, as a function of time:
\begin{equation}\label{2} R(t)=R_S\left[1-e_{free}\cos \left(v_et+\kappa\right)+\sum_{k=1}^\infty C_k \cos \left(k n_{syn}t\right)\right]\,,
\end{equation}
\begin{equation}\label{3} 
z(t)=iR_S\cos \left(v_it+\lambda\right)\,.
\end{equation}
Here $n_{syn}$ stands for the synodic frequency, i.e. $n_{syn}=n_{PC}-n_S$ and $i$ is the inclination of the moon orbit with respect to the Pluto-Charon orbital plane. The moon's mean motion $n_{S}$, the epicyclic frequency $v_e$ and the vertical frequency $v_i$ are defined as follows (see Appendix in \cite{Bromley-Kenyon:2020}):
\begin{eqnarray}\label{4} &n_{S}^2=\frac{1}{R_S}\frac{d\Phi _{00}}{dR}\Bigg|_{R_S}  =\frac{GM}{R_S^3} \Bigg\{ 1+ \frac{\mu}{M} \nonumber \\ &\times \Bigg[ \frac{3}{4} \frac{a_{PC}^2}{R_S^2} + \frac{45}{64} \frac{\mathcal{M}_3^{(+)}}{M^3} \frac{a_{PC}^4}{R_S^4} + \frac{175}{256} \frac{\mathcal{M}_5^{(+)}}{M^5} \frac{a_{PC}^6}{R_S^6} + {\cal O}\left(\frac{a_{PC}^8}{R_S^8}\right) \Bigg] \Bigg\} \end{eqnarray}
\begin{eqnarray}\label{5} 
&v_{e}^2=R_S\frac{dn _{S}^2}{dR}\Bigg|_{R_S}+4n _{S}^2= \frac{GM}{R_S^3} \Bigg\{ 1 - \frac{\mu}{M} \nonumber \\
&\times \Bigg[\frac{3}{4} \frac{a_{PC}^2}{R_S^2} + \frac{135}{64} \frac{\mathcal{M}_3^{(+)}}{M^3} \frac{a_{PC}^4}{R_S^4} + \frac{875}{256} \frac{\mathcal{M}_5^{(+)}}{M^5} \frac{a_{PC}^6}{R_S^6} + {\cal O}\left(\frac{a_{PC}^8}{R_S^8} \right)\Bigg] \Bigg\} \end{eqnarray}
\begin{eqnarray}\label{6} 
&v_{i}^2=\frac{1}{z}\frac{d\Phi }{dz}\Bigg|_{z=0,\, R_S}= \frac{GM}{R_S^3} \Bigg\{ 1 - \frac{\mu}{M} \nonumber \\
&\times \Bigg[\frac{9}{4} \frac{a_{PC}^2}{R_S^2} + \frac{225}{64} \frac{\mathcal{M}_3^{(+)}}{M^3} \frac{a_{PC}^4}{R_S^4} + \frac{1225}{256} \frac{\mathcal{M}_5^{(+)}}{M^5} \frac{a_{PC}^6}{R_S^6} + {\cal O}\left(\frac{a_{PC}^8}{R_S^8}\right) \Bigg] \Bigg\} \end{eqnarray}
where $M=M_P+M_C$ and $\mu=(M_P \cdot M_C)/(M_P+M_C)$ are the total and reduced mass of the binary, respectively, and $\mathcal{M}_a^{(+)}=M_P^a + M_C^a$ (the subscripts denote the respective objects). 

The factor $C_k$ represents the amplitudes of the oscillations:
\begin{equation} \label{8}
C_k=\left[\frac{1}{R_S}\frac{d\Phi _{k}}{dR}\Bigg|_{R_S}-\frac{2n_S\Phi_k}{R_S^2n_{syn}}\right]\frac{1}{v_e^2-k^2n_{syn}^2}
\end{equation}
Assuming that the whole system’s barycenter coincides with the Pluto-Charon’s barycenter, the orbital distance of a small moon would be: 

\begin{equation}\label{10} r(t)=\sqrt{R^2(t)+z^2(t)}\end{equation}
%


We assign the nominal eccentricities (Table~\ref{tab:1}) of each moon as $e_{free}$. Table~\ref{tab:2} presents the calculated values of the major frequencies at which the moons oscillate. The central binary frequency is 0.1566 days$^{-1}$ (according to Table~\ref{tab:1}). 

\begin{table}
	\centering
	\caption{Orbital parameters for Pluto’s moon system. The semi-major axes are given with respect to the system's center of mass. The masses are based on the results of \protect\cite{Buie:2012} and all the other  parameters are adopted from \protect\cite{Showalter:2015}.}
	\label{tab:1}
	\begin{tabular}{lccc} 
		\hline
		Object & Mass ($10^{16}$ kg) & Semi-major axis (km) \\
		\hline
		Pluto & 1,303,000 & 2,126  \\
Charon & 158,600 & 17,470  \\
Styx & 0.06 & 42,656 $\pm$ 78  \\
Nix & 4.5 & 48,694 $\pm$ 3 \\
Kerberos & 0.1 & 57,783 $\pm$ 19 \\
Hydra & 4.8 & 64,738 $\pm$ 3 \\
		\hline
        \end{tabular}
        
        \begin{tabular}{ccc}
        \hline
	Period (days) & Eccentricity  ($10^{-3}$) & Inclination ($^{\circ}$)\\

		\hline
	6.3872273 & 0.000 & 0.000 \\
6.3872273 & 0.000 & 0.000\\
20.16155 $\pm$ 0.00027 & 5.787 $\pm$ 1.144 & 0.809 $\pm$ 0.162 \\
24.85463 $\pm$ 0.00003 &  2.036 $\pm$ 0.050 & 0.133 $\pm$ 0.008 \\
32.16756 $\pm$ 0.00014 & 3.280 $\pm$ 0.200 & 0.389 $\pm$ 0.037 \\
38.20177 $\pm$ 0.00003 & 5.862 $\pm$ 0.025 & 0.242 $\pm$ 0.005 \\
		\hline
	\end{tabular}
\end{table}

\begin{figure}
	a\includegraphics[width=\columnwidth]{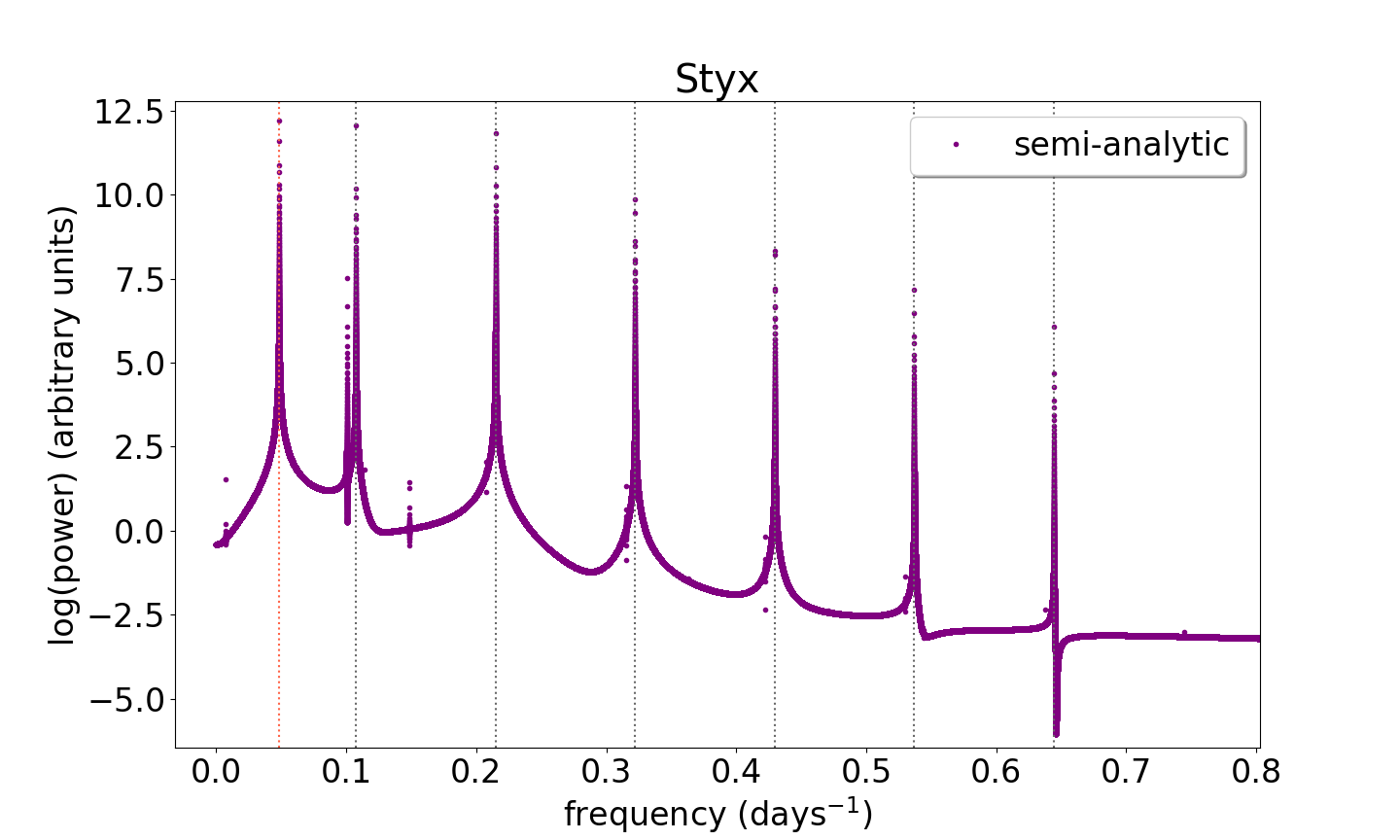}\\
    b\includegraphics[width=\columnwidth]{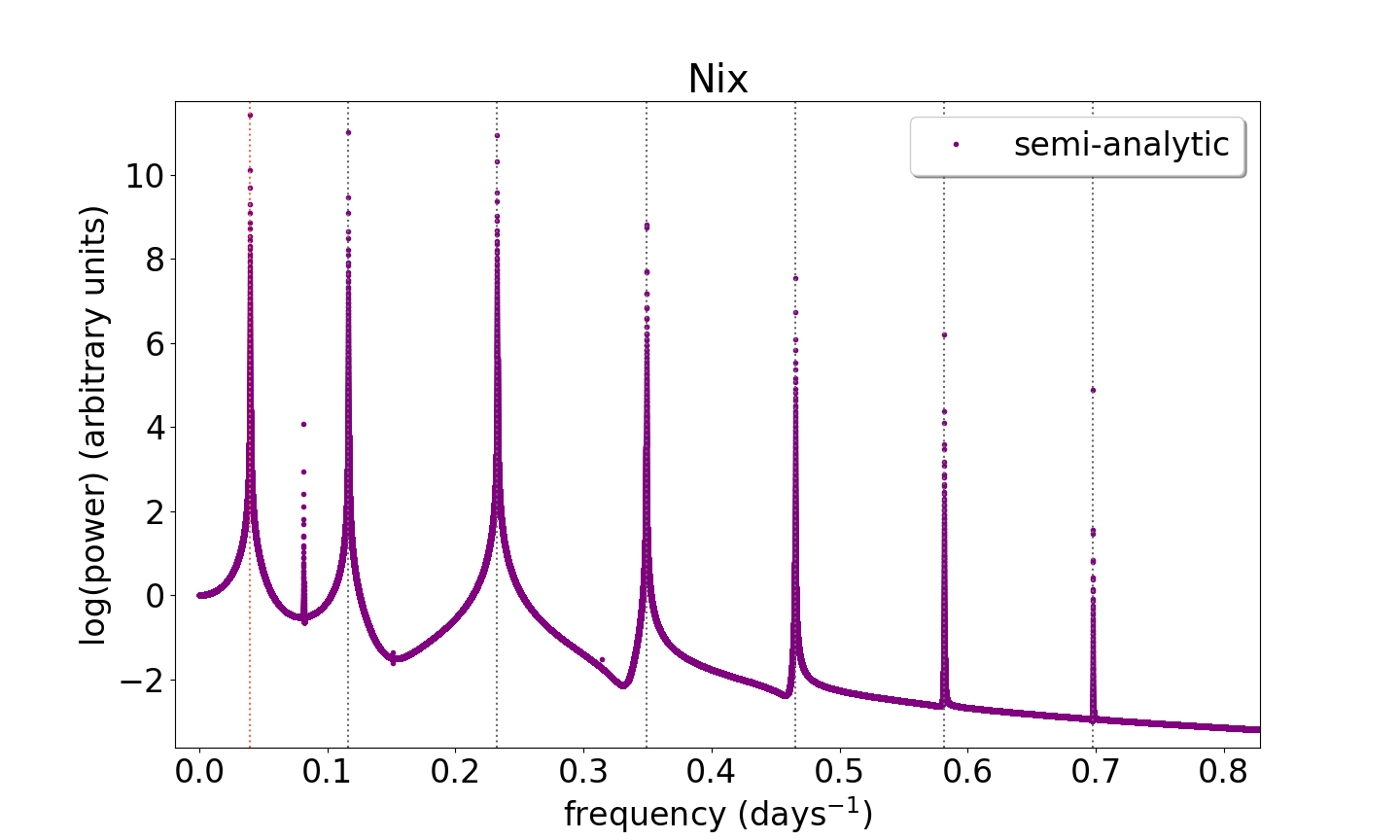}\\
    c\includegraphics[width=\columnwidth]{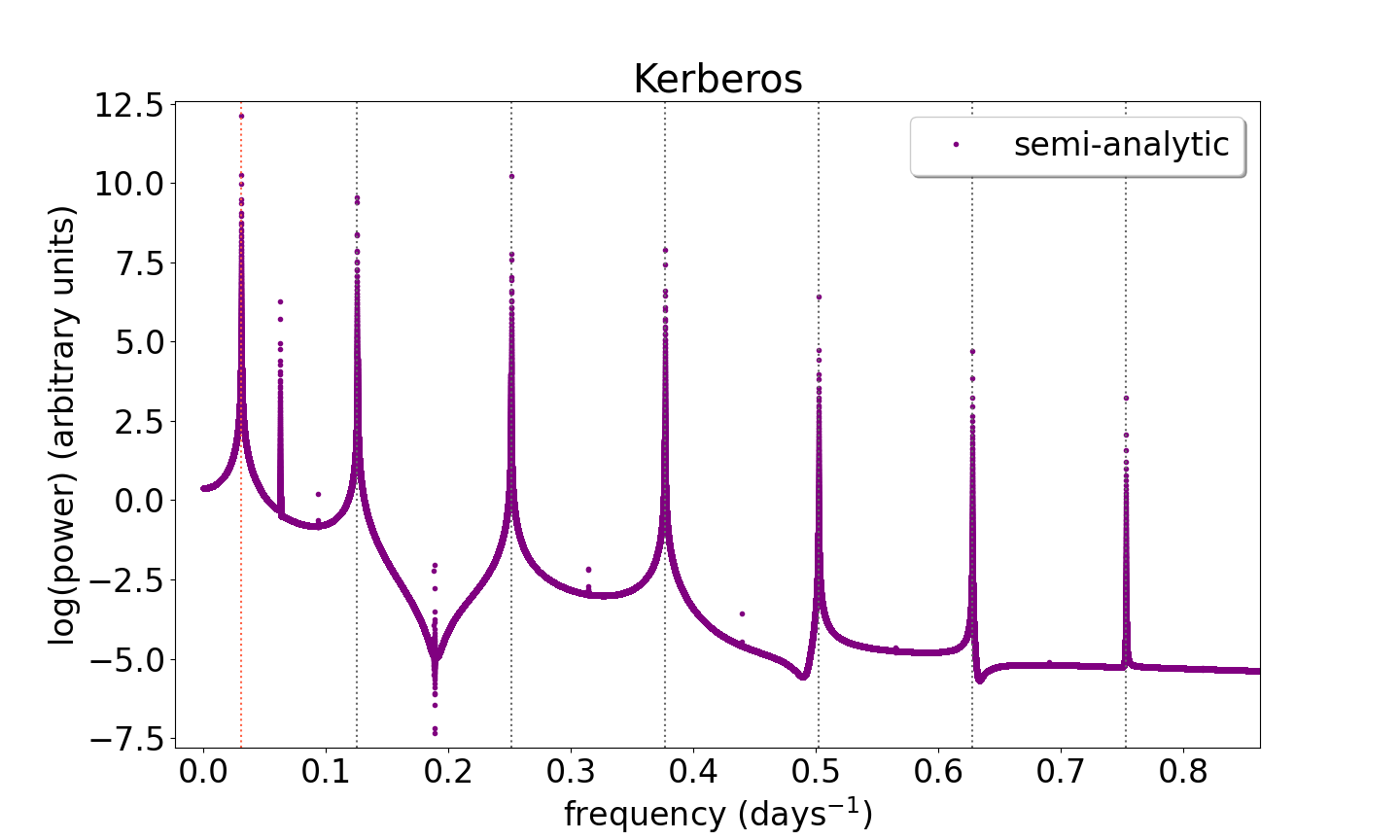}\\
    d\includegraphics[width=\columnwidth]{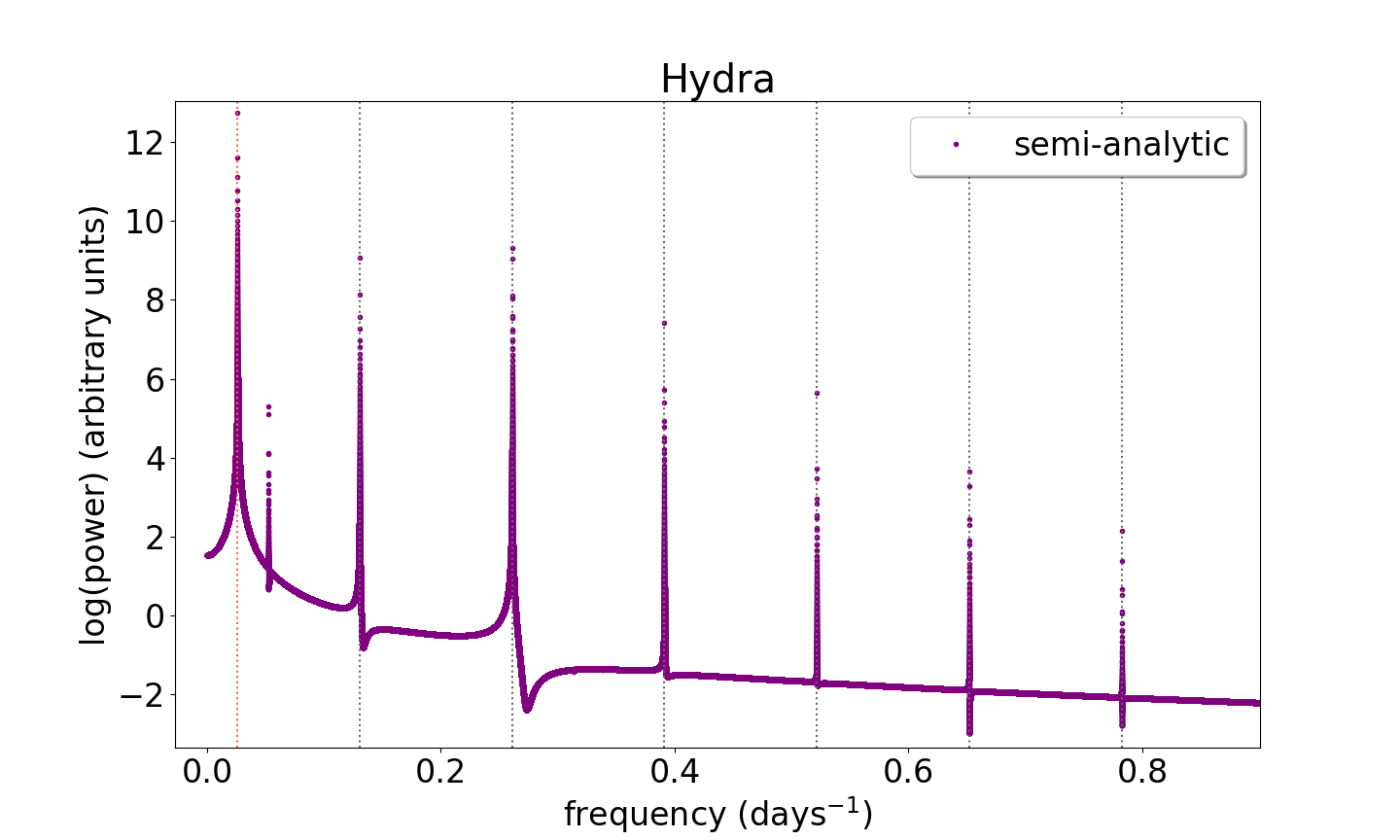}
    \caption{FFT power spectrum for all small moons using the semi-analytic approach. In each panel the red vertical line represents the epicyclic frequency of the moon ($v_e$) and the gray lines the synodic frequency and its harmonics ($k\, n_{syn}$), as shown in Table~\ref{tab:2}.}
    \label{fig:1}
\end{figure}

\subsection{Numerical approach}
\label{sec:2.2}

We approach the orbits numerically by implementing an n-body symplectic integrator in a Python 3.9.6 IDLE environment. The n-body simulation code 
utilizes the kick-drift technique to solve the differential equations representing gravitational interactions. The desired accuracy of the code is validated, since the total calculated energy of the system is kept constant. 

Simulations of the 6-body system in question, with different initial data sets each time, have been performed using the same n-body code by \cite{Gakis:2022}. We re-examine the basic situation of these simulations, in order to give a more thorough perspective on the concepts discussed in this paper. Specifically, initial conditions are taken from \cite{Brozovic:2015}, where a table is provided (Table 8 therein) of measured 3D vectors of positions and velocities for every object. We analyzed the evolution of the system forward in time using this data set.

The numerical integration timestep is fixed to $\Delta t=5000\,s $, which maintains computational times under manageable limits, and at the same time keeps uncertainties below 0.1\%, as determined in \cite{Gakis:2022}. Besides, \cite{Kenyon:2019b,Kenyon:2019a} propose at least $\Delta t \lessapprox13,500\,s$ for reliable integrations. Timesteps like these are used by various studies on the orbits within the Pluto-Charon system; for example the numerical calculations performed by \cite{Woo:2020} have a $\Delta t=3,000\,s $, whereas \cite{Lee:2006} use a larger timestep, $\Delta t=10,000\,s $. Nevertheless, we also ran numerical tests with smaller timesteps, not identifying however any 
noticeable changes. The gravitational effect of the Sun is significant at distances of about ten times larger than Hydra's average orbital distance \citep{Michaely:2017}; therefore, we focus on a restricted 6-body problem in our simulations, neglecting the Sun and other Solar System bodies.

\begin{figure}
	a\includegraphics[width=\columnwidth]{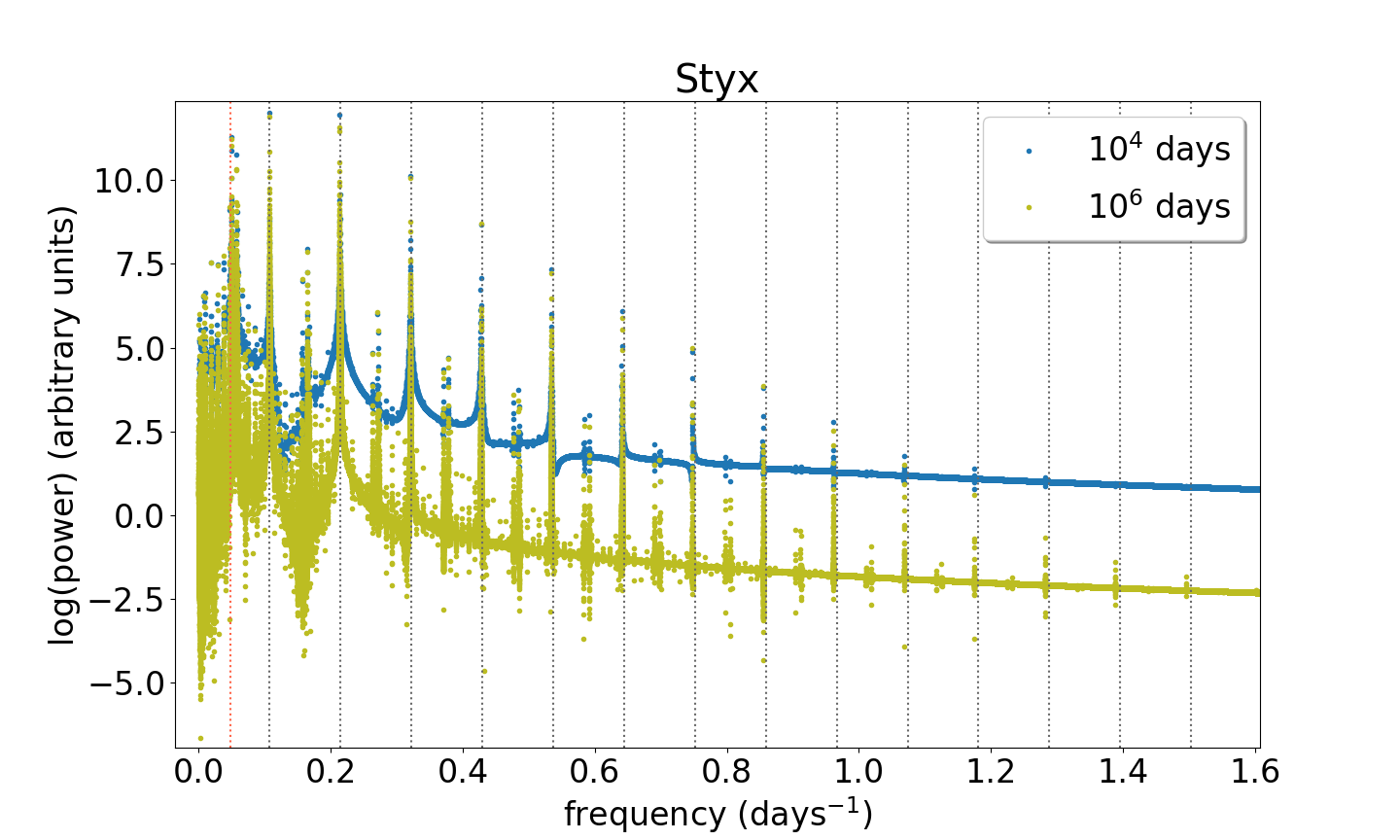}\\
    b\includegraphics[width=\columnwidth]{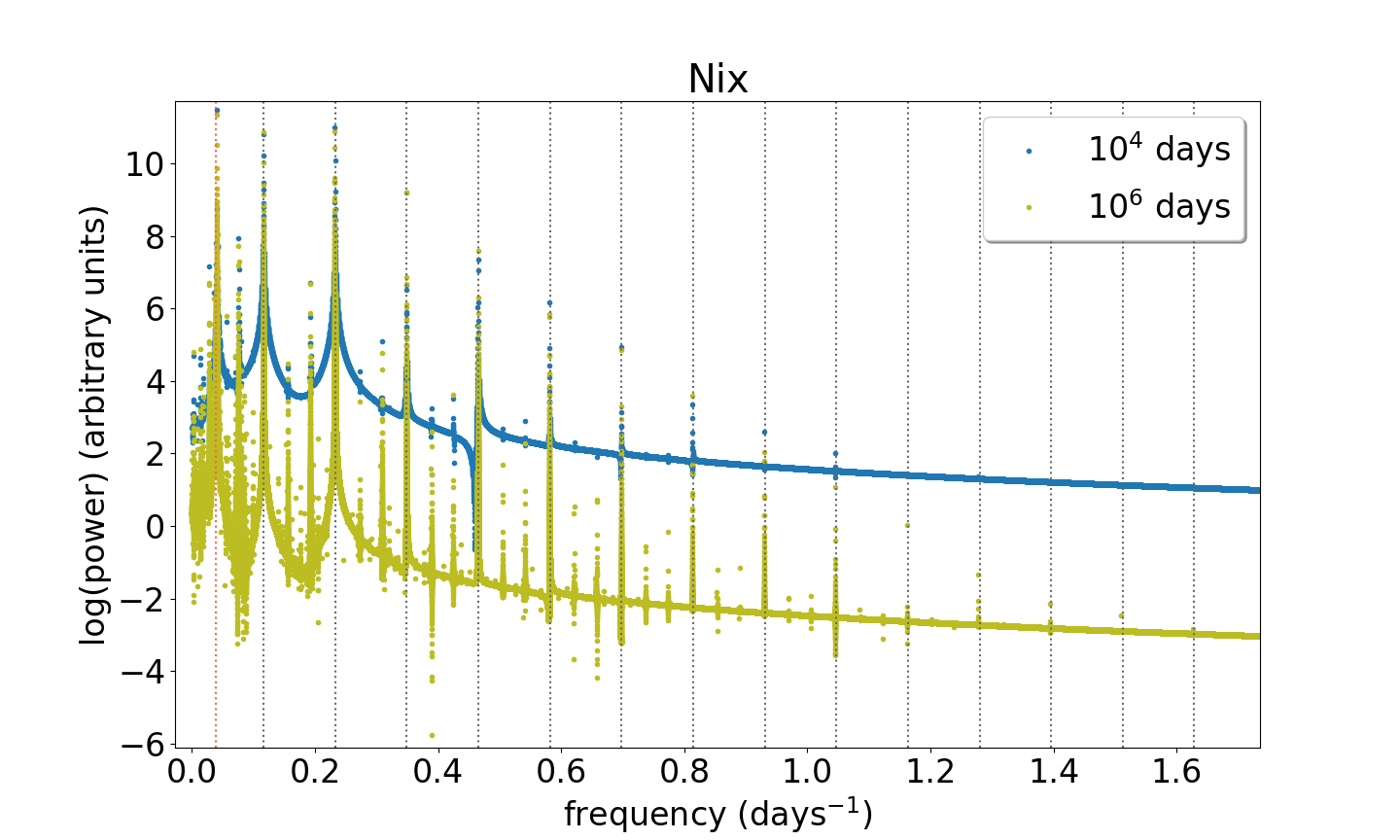}\\
    c\includegraphics[width=\columnwidth]{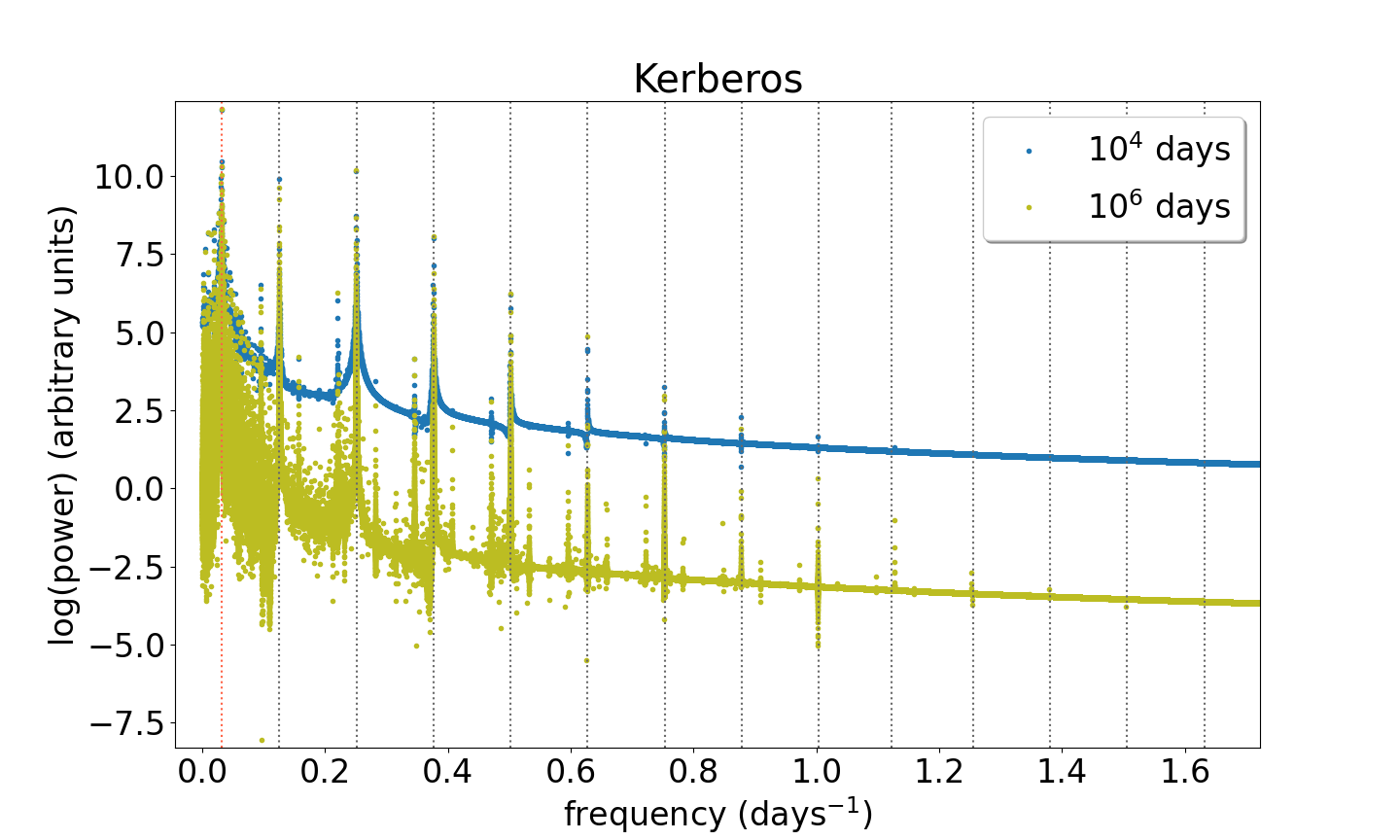}\\
    d\includegraphics[width=\columnwidth]{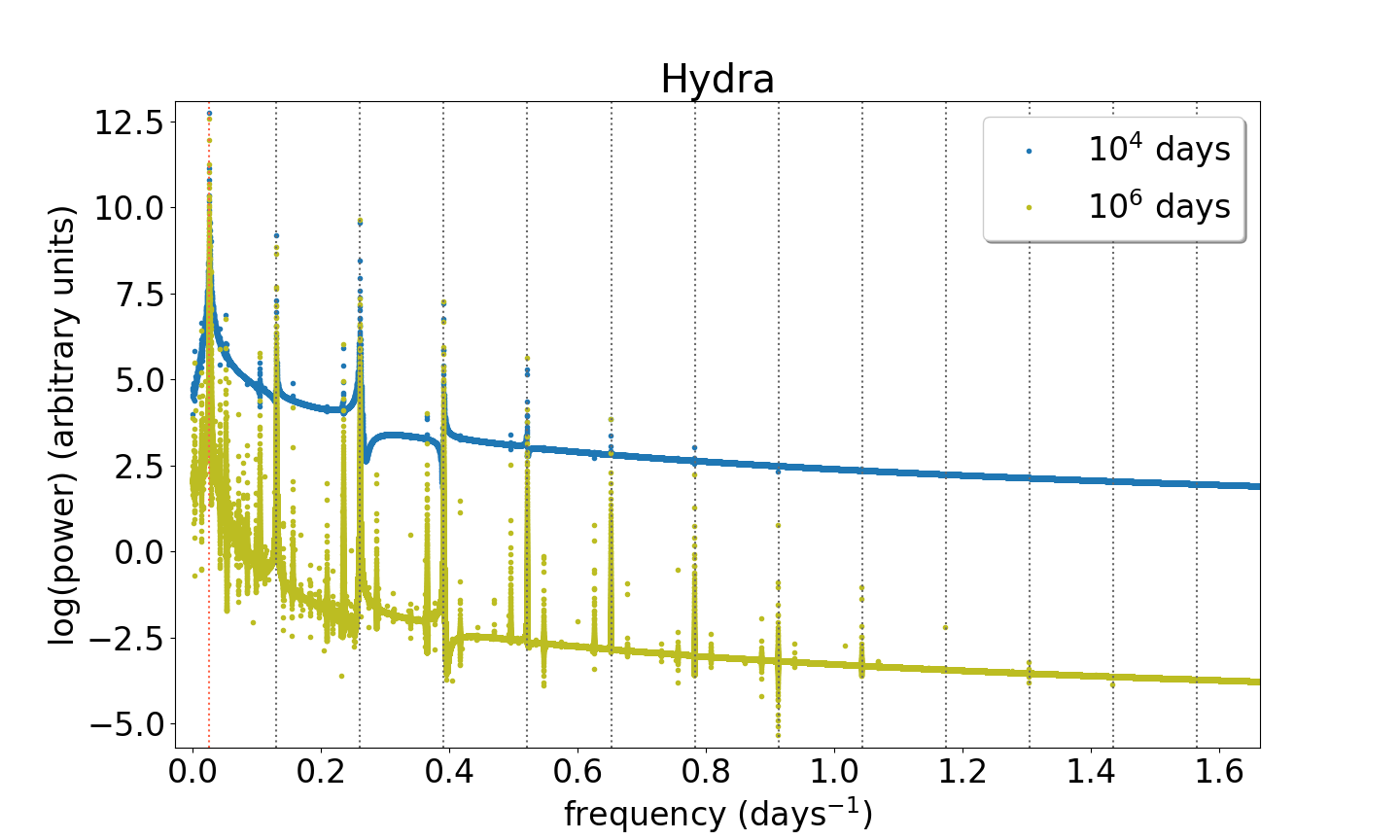}
    \caption{FFT power spectrum for all small moons by varying the total simulated time of 6-body integrations (10$^4$ and 10$^6$ days). In each panel the red vertical dotted line represents the epicyclic frequency of the moon ($v_e$) and the gray lines the synodic frequency and its harmonics ($k\,n_{syn}$), as shown in Table~\ref{tab:2}.}
    \label{fig:2}
\end{figure}

\begin{figure}
	a\includegraphics[width=\columnwidth]{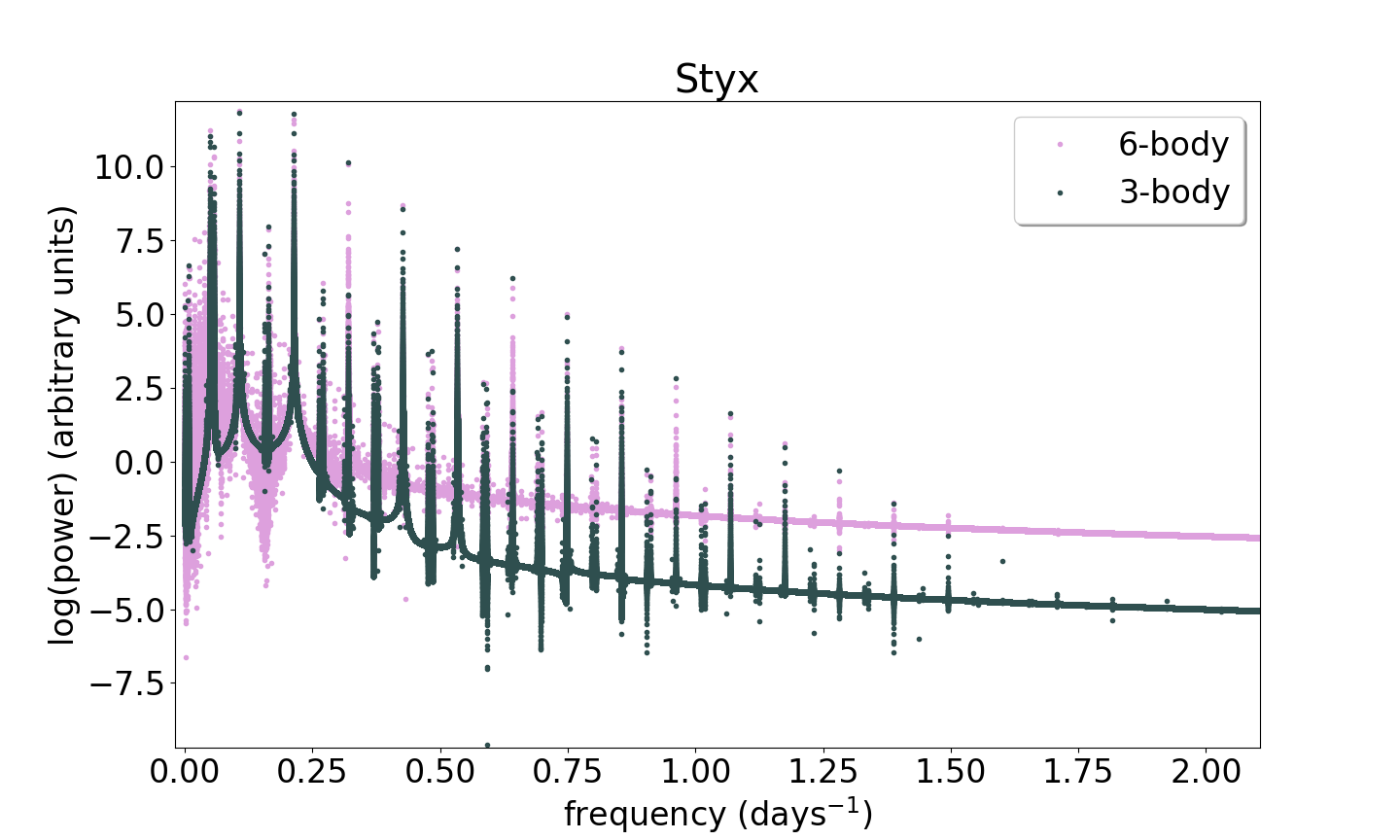}\\
    b\includegraphics[width=\columnwidth]{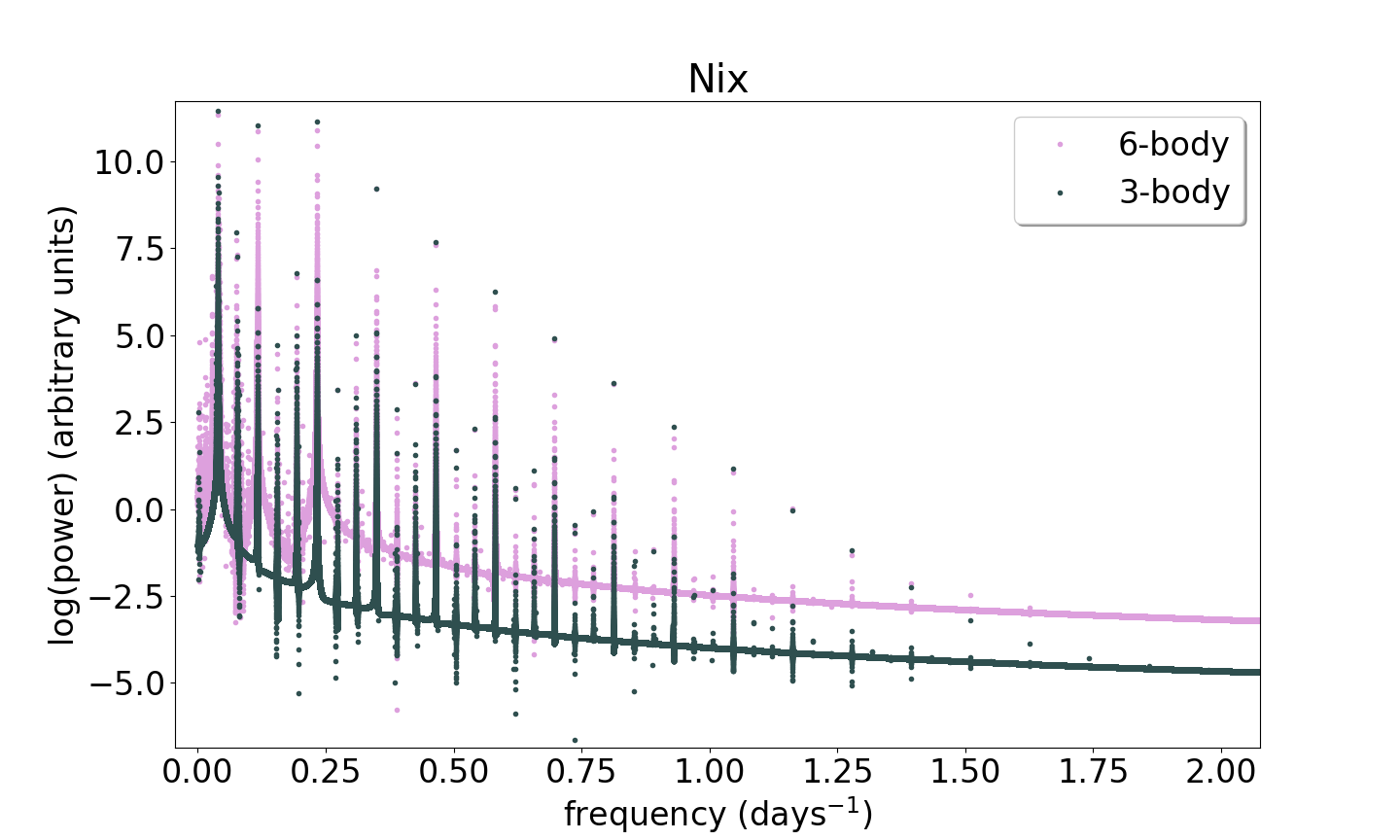}\\
    c\includegraphics[width=\columnwidth]{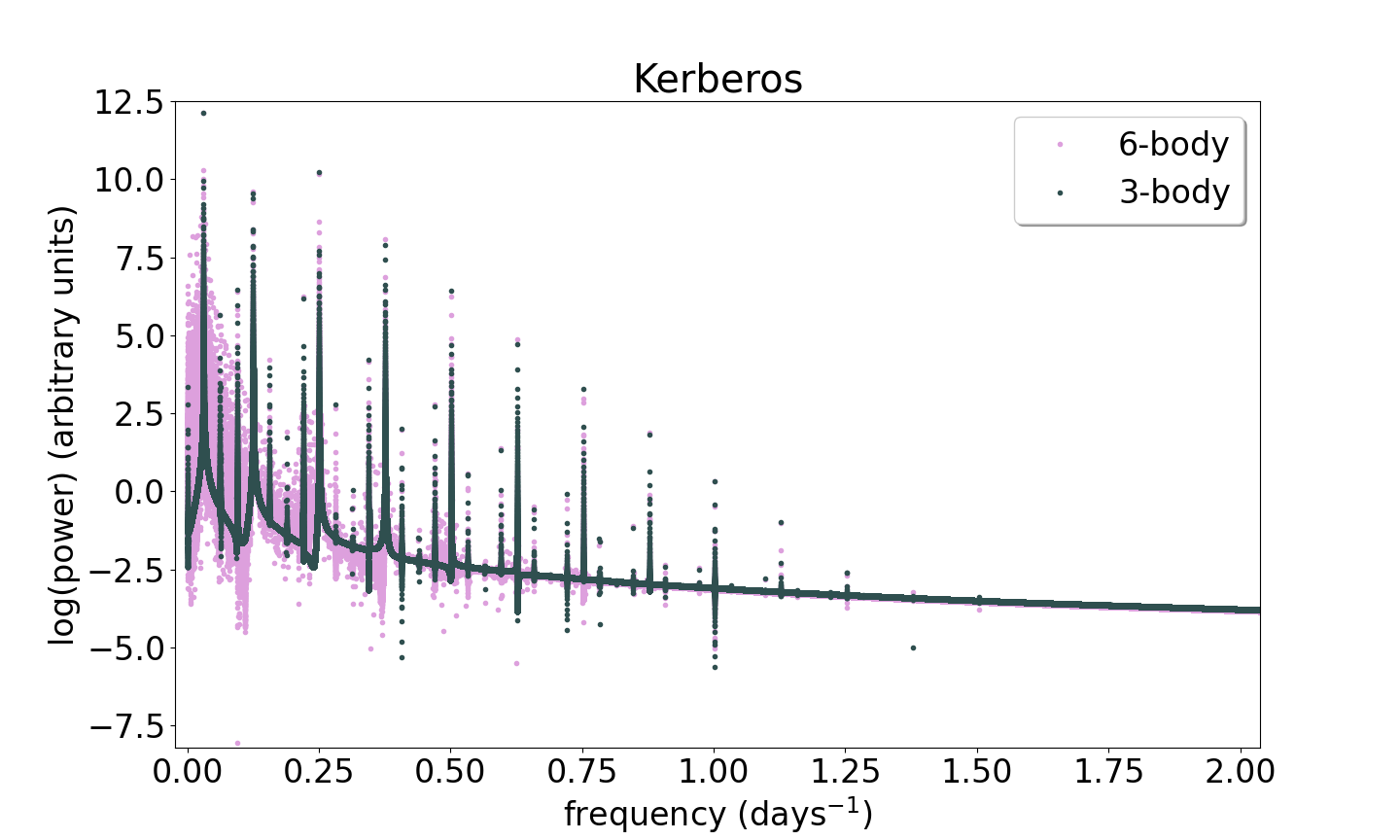}\\
    d\includegraphics[width=\columnwidth]{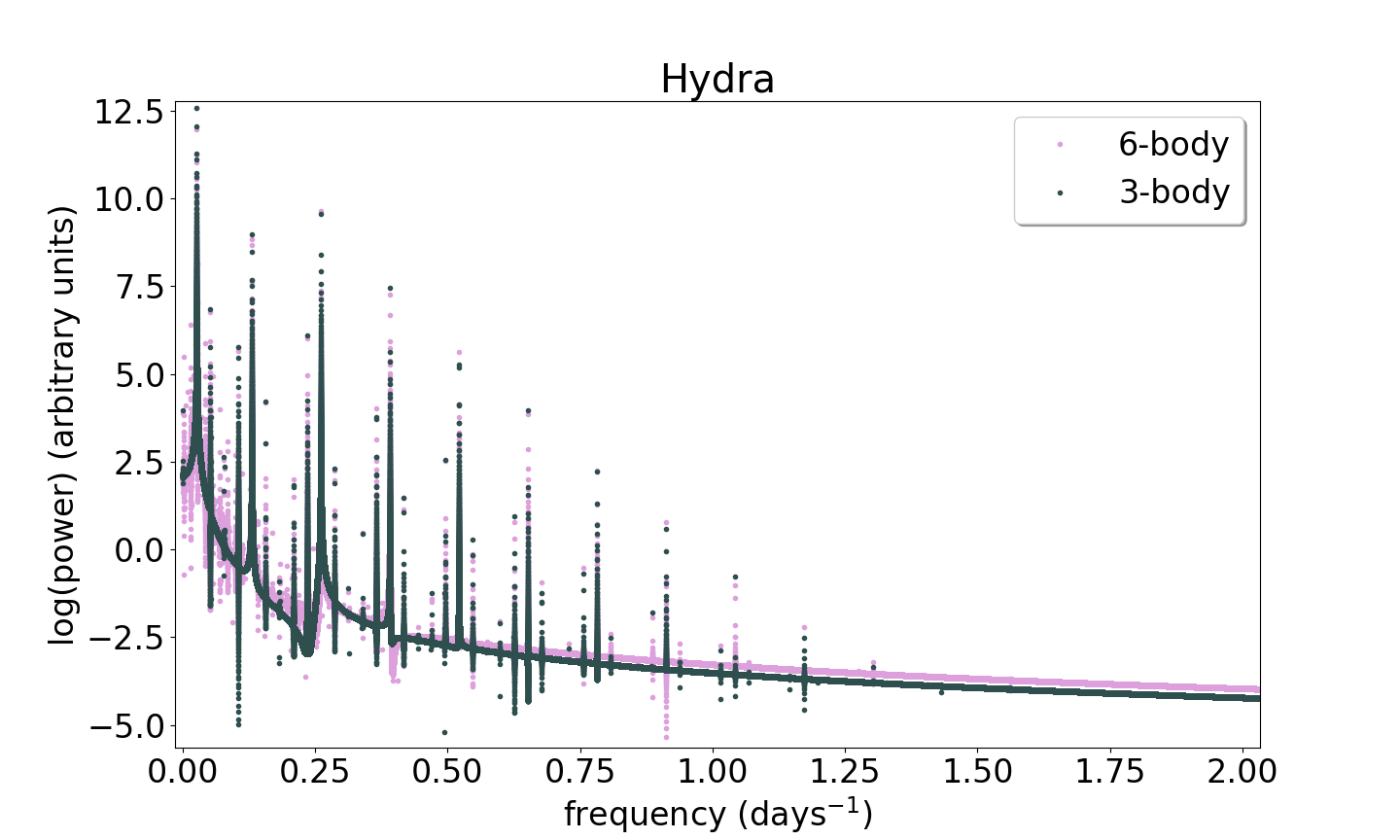}
    \caption{FFT power spectrum for all small moons for 6-body (violet) and
3-body integrations (dark blue). The simulated time is 10$^6$ days ($\sim$ 2700 years).}
    \label{fig:3}
\end{figure}

\section{Results}
\label{sec:3} 

\begin{table}
	\caption{The first major oscillatory frequencies, in ($2\pi$ days)$^{-1}$, for all small moons. The rest are calculated similarly.}
	\label{tab:2}
        \begin{tabular}{lllll}
        \hline\noalign{\smallskip}
       Parameter & Styx & Nix & Kerberos & Hydra \\
		\noalign{\smallskip}\hline\noalign{\smallskip}
		$n_S$ & 0.0492 & 0.0402 & 0.0311 & 0.0262 \\
        $v_e$ & 0.0482 & 0.0396 & 0.0308 & 0.0260\\
        $v_i$ & 0.0502 & 0.0408 & 0.0314 & 0.0264\\
$n_{syn}$ & 0.1074 & 0.1163 & 0.1255 & 0.1304\\
		\noalign{\smallskip}\hline
	\end{tabular}
\end{table}

Several algorithms are implemented in order to study the orbits in several dynamical systems. In our analysis, we chose to adopt FFT to decompose the orbits in the Pluto-Charon system. Although FFT may be less accurate than other methods, such as the Frequency Map Analysis (FMA) \citep{Laskar:1999}, it still produces reliable outcomes significantly quickly, and hence is often applied to analyze circumbinary orbits \cite[e.g.][]{Woo:2020,Gakis:2022bb}. Our results indicate that the resolution provided by FFT is suitable to detect and separate the forced frequencies caused by the central binary and some of their harmonics, as well as the main trends of the reciprocal effects by each of the other moons. Besides, since our goal is to identify the frequencies of the moon's oscillations rather than studying chaos in the system, which has been explored thoroughly in past studies \cite[e.g.][]{Kenyon:2019b,Kenyon:2019a,2022AJ....163..238K}, the FFT method will suffice.

The general formula used to 
convert a sequence $x[n] $ of length $N$ into a new one $y[k]$ using a Fourier transform is:

\begin{equation}\label{11} y[k]=\sum_{n=0}^{N-1}e^{-2\pi j \frac{kn}{N}}x[n]\end{equation}
More precisely, we convert a distance domain into a domain of frequencies. 
Fast Fourier transforms are 
performed using the Python \texttt{scipy} routine \texttt{fft}.

\subsection{Central binary effects}
\label{sec:3.1} 

At first, we apply FFT of $r(t)$ for the outcomes of the semi-analytic model (Fig.~\ref{fig:1}). The timestep adopted for the semi-analytic calculations was the same with the n-body integration timestep, and the total duration was set to 10$^6$ days. Unlike \cite{Woo:2020}, here we examine the vertical motion, since our calculations include the proper inclinations. The most outstanding frequencies arising are those defined when computing Eqs.~(\ref{2}) and~(\ref{3}), as expected. The red vertical dotted line in each periodogram corresponds to the value of $v_e$, whereas the gray lines correspond to the harmonics $k(n_{PC}-n_S)$. The spikes vary in height, as anticipated by the factor $C_k$. In other words, the relative size of each peak would give us a comparison of the different amplitudes of each frequency. 

As far as the vertical frequency $v_i$ is concerned, there also appears to be a minuscule peak, though not visible in the frequency spectra of Fig.~\ref{fig:1}. Having zoomed into the low-frequency area of each spectrum and identifying a corresponding (barely visible) formation, we advocate that its apparent absence is not a problem of the frequency resolution nor with the wide frequency range. Instead, this is a result of the factor $i\,R_S$ in equation~(\ref{3}) outlining the vertical frequency, which much smaller than the values of $R_S$ in equation~(\ref{2}), which dominates in the configuration of the strength of each peak in the frequency spectrum ($i < 1^{\circ}$  for all moons). 

There are also some other secondary spikes, not corresponding to the values of Table~\ref{tab:2}. They originate from the vertical component of the motion and the sinusoidal products deriving from equation~(\ref{10}). Namely, $R^2(t)$ and $z^2(t)$, along with the square root, give a number of harmonic cross terms that eventually result in frequencies of forms like $v_e \pm k\,n_{syn}$ (and multiples), $2v_e$, $2v_i$ and so on. In general, many combinations of $v_e$, $k\,n_{syn}$ and $v_i$ arise from equation~(\ref{10}), which are present in the periodograms of Fig.~\ref{fig:1}. Most of the secondary peaks have a quite small amplitude, which makes them only clearly visible when using a highly magnified image. 

There are some distinctive differences once the n-body simulation is employed. The resulting power spectra are shown in Fig.~\ref{fig:2}. The system is let to evolve for 10$^4$ and 10$^6$ days. Again, in this figure, vertical lines show the main expected frequencies, as they have been computed in Table~\ref{tab:2}. The primary peaks are observed at these frequencies in this case as well. Nevertheless, a large number of other minor peaks are also visible. In fact, when we increase the simulated time, additional frequencies appear or, alternatively, the already-present frequencies are enhanced. Furthermore, the increase of the total timespan reduces unwanted noise; the values of the power spectrum are diminished. 

We notice that the vertical lines (i.e. the findings of the semi-analytic model) do not exactly match with the peaks of the periodogram in Fig.~\ref{fig:2}. This is not largely visible on large scale (deviations scale to $\sim$0.5\%), but may be observed when zooming in on each individual spike. This distinction is evidence of the unavoidable inconsistency between the two approaches that we follow. Any differences could safely be attributed to approximations made in the semi-analytic model, as justified in Section \ref{sec:2}. For example, we neglect the nonlinear terms that definitely rule the motions of the moons. Styx is the most striking example and reveals the limitations of the \cite{Lee:2006} theory in distances close to the binary.

Yet, apart from the peaks dominating the region around zero, which we discuss later, the low-amplitude spikes appearing in between the vertical lines can clearly be understood within the semi-analytic model. As we discussed earlier for Fig.~\ref{fig:1}, these peaks are the result of the multiple sinusoidal products of equation~(\ref{10}). Thus, they are not in principal caused by numerical errors, but are undeniably anticipated by the theoretical model.

\cite{Bromley-Kenyon:2020} and \cite{Woo:2020} found further formations in their power spectra, unable to be explained by the epicyclic theory. In the first paper, the authors found a residual signal at the epicyclic frequency, when simulating a most-circular orbit for Nix (purple curve in their Figure 2). We believe that attributing this residual solely to numerical challenges seems unlikely, as it occurs in the exact frequency of $v_e$ and appears prominently when increasing $e_{free}$. We argue, instead, that this behavior most probably reflects the practical impossibility of defining a zero-eccentric circumbinary orbit, as validated by \cite{Gakis:2022}. In other words, although the linearized theory allows an orbit exempt of free eccentricity, even adopting such initial conditions inevitably yields that some frequencies will couple to $v_e$.  Accordingly, forced frequencies of the form $n_{syn}-v_e$ are expected, as we explained earlier. In the latter study, we inspect that peaks near the values of $n_S$ and $v_e$ in Figures 5 and 6 of \cite{Woo:2020} might truly appear at some extent from higher-order terms, but perhaps are generated by the merging of the frequencies $v_e$, $k\,(n_{PC}-n_S)$, and $v_i$, as quantified by equation~(\ref{10}).

\subsection{Mutual interactions}
\label{sec:3.2} 

\begin{figure*}
\begin{tabular}{p{0.49\textwidth} p{0.5\textwidth}}
  a\includegraphics[width=0.49\textwidth]{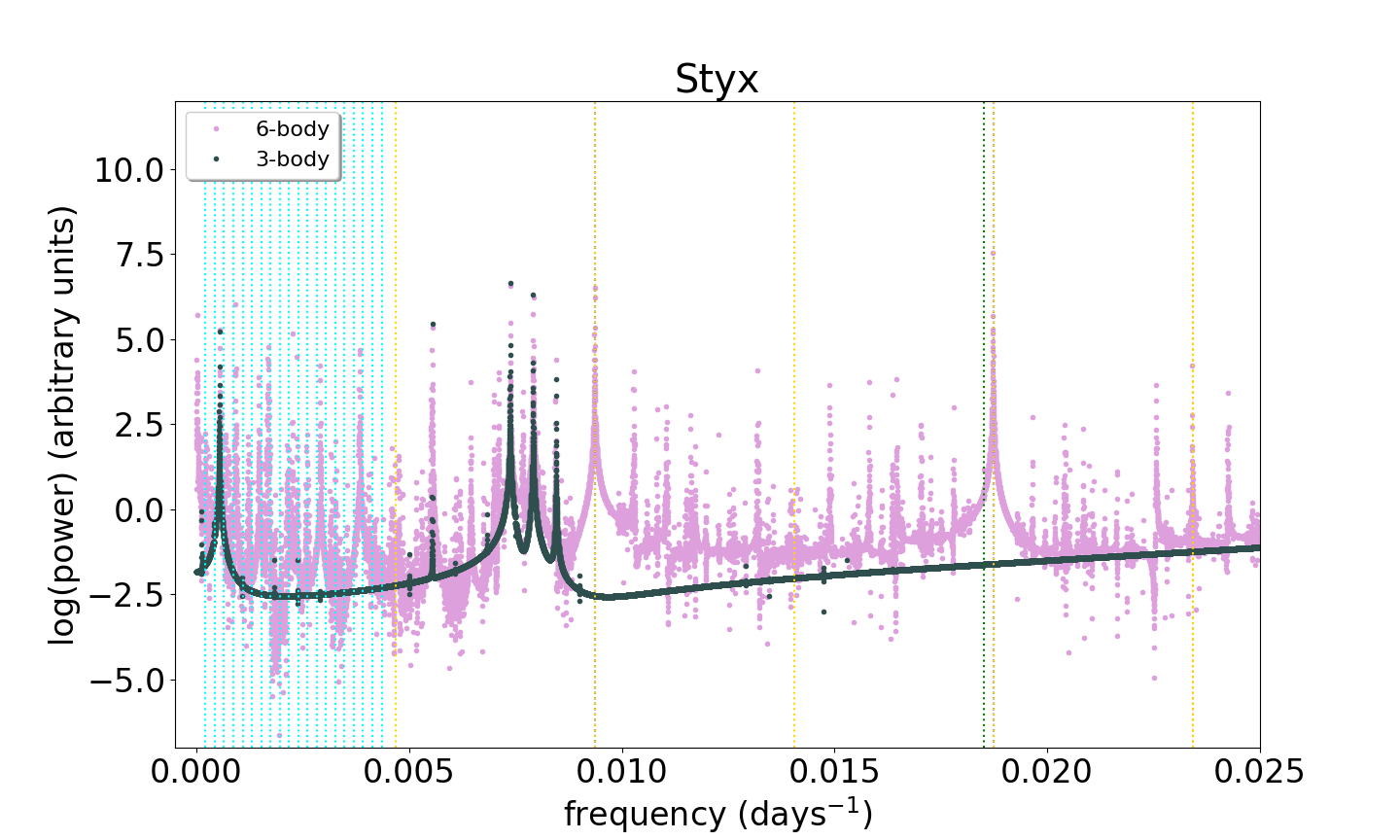}&
    b\includegraphics[width=0.49\textwidth]{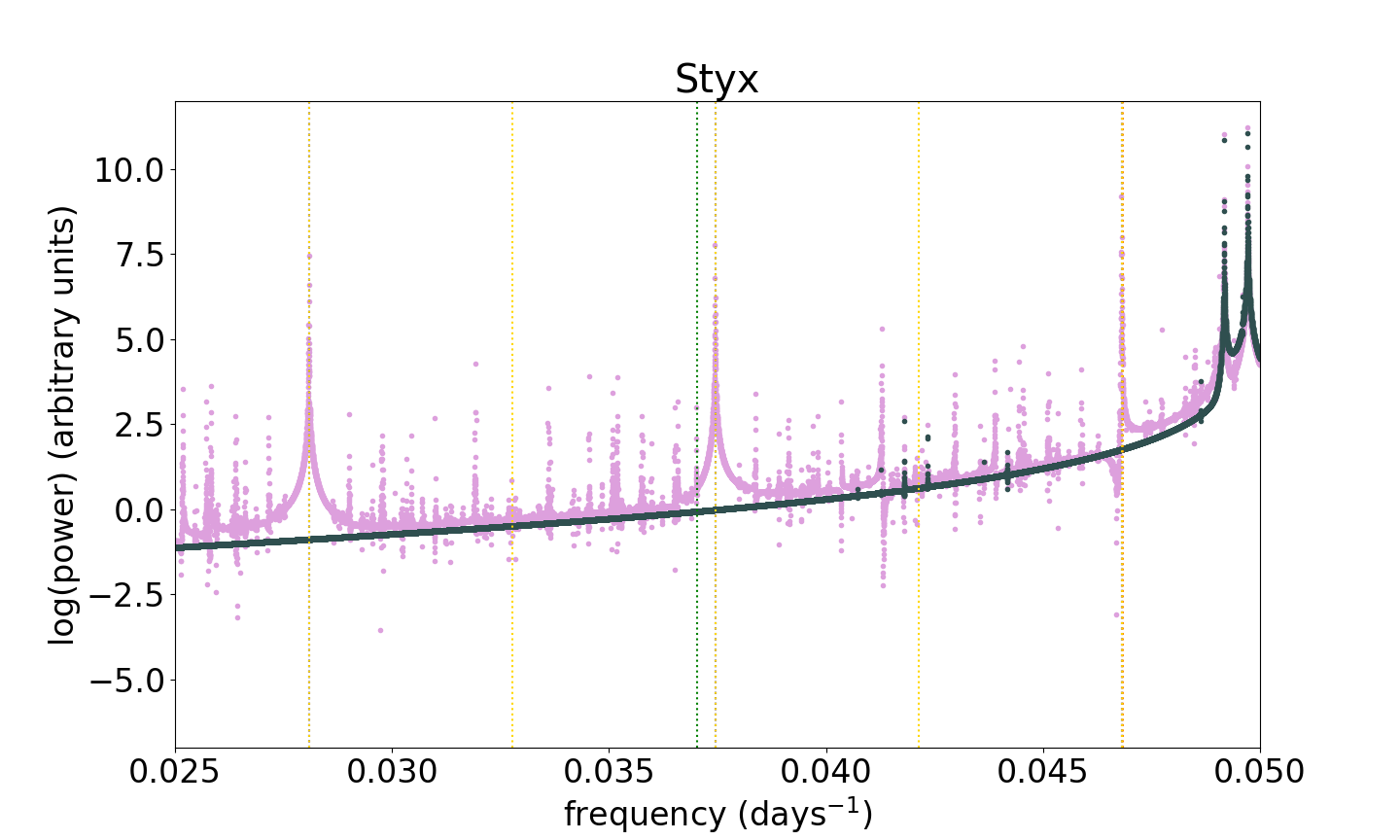} \\
     c\includegraphics[width=0.49\textwidth]{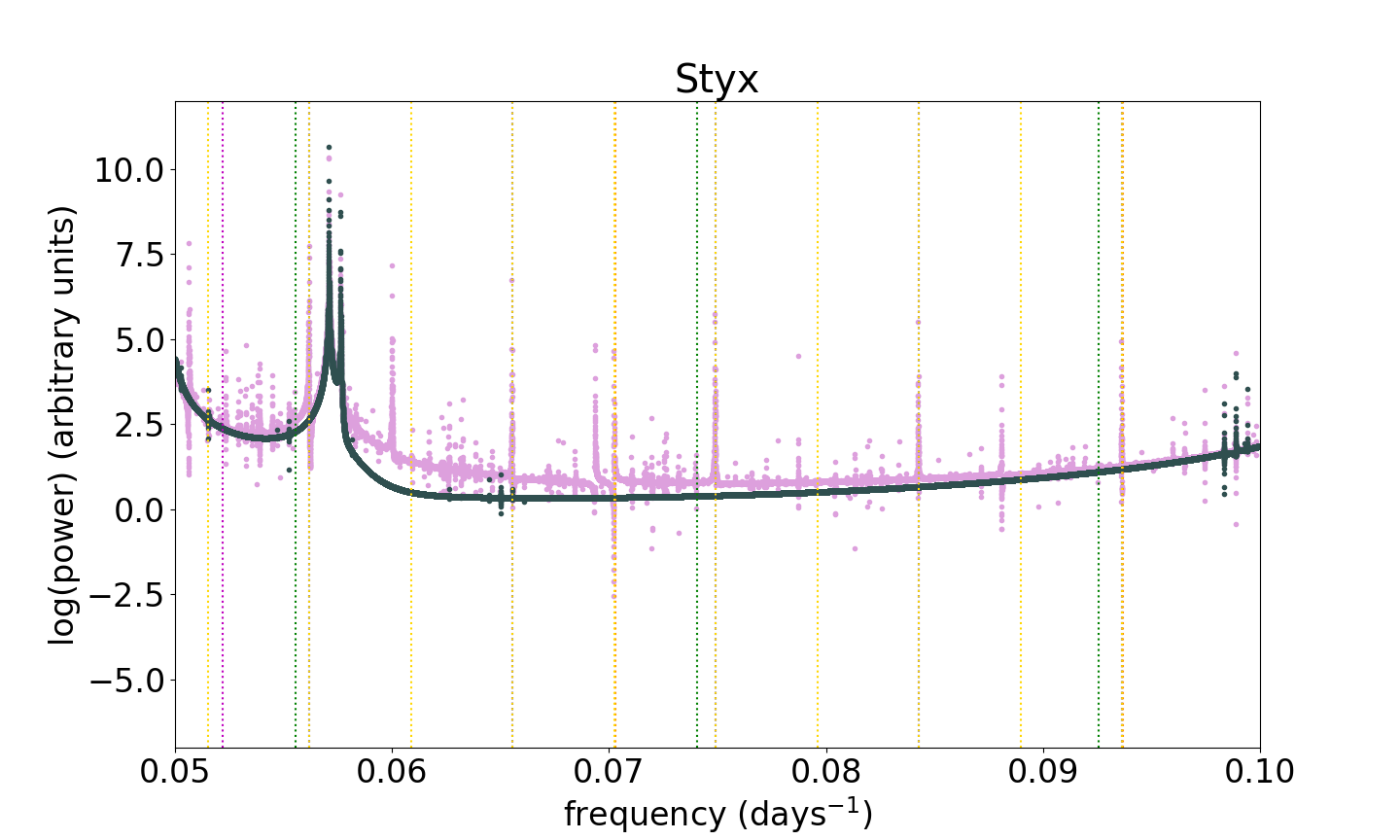} &
       d\includegraphics[width=0.49\textwidth]{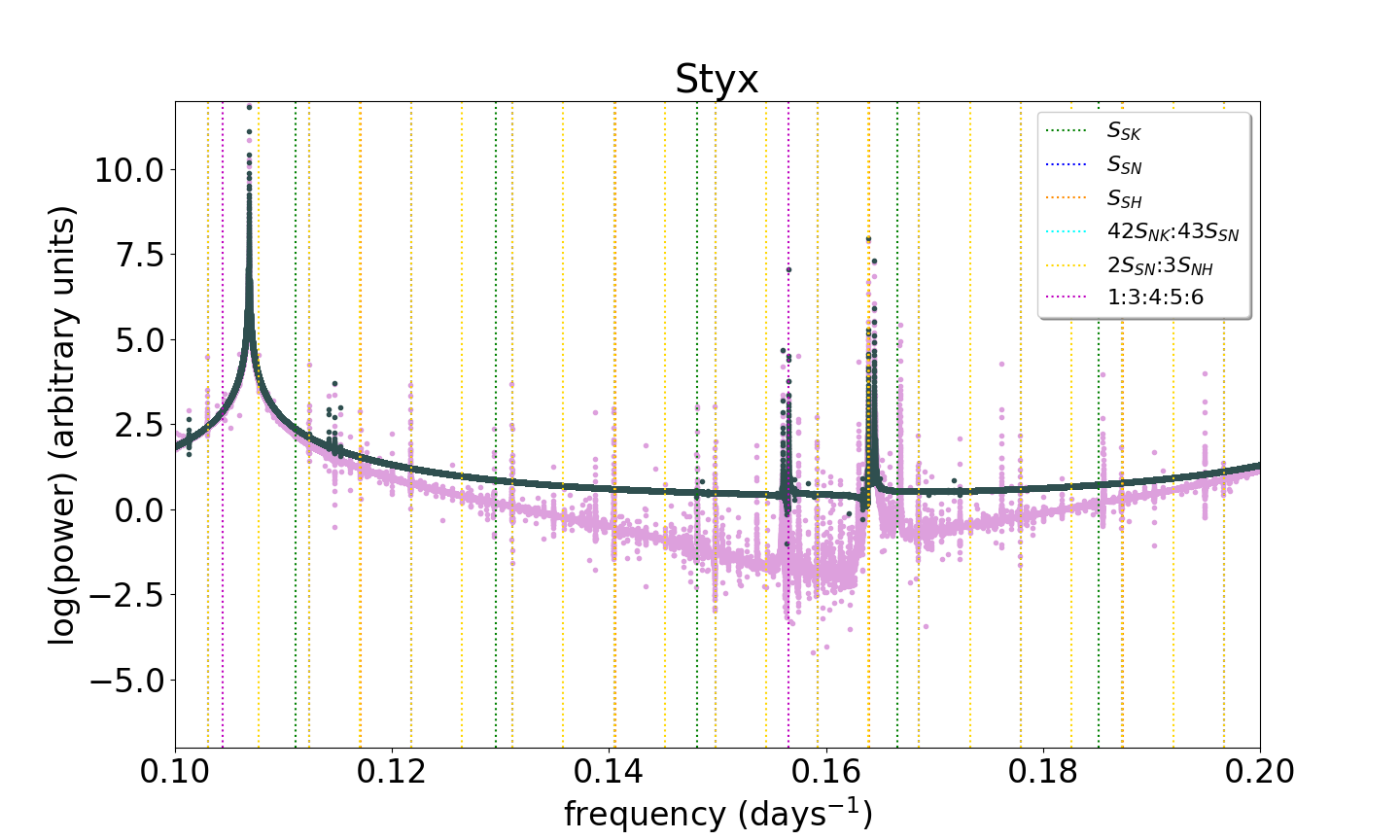} \\
     \end{tabular}
\caption{FFT successive power spectra for Styx, magnified in low frequencies. Panel a includes frequencies $\leq$ 0.0025 days$^{-1}$, panel b between 0.0025 days$^{-1}$ and 0.050 days$^{-1}$, panel c between 0.050 days$^{-1}$ and 0.10 days$^{-1}$ and panel d $\geq$ 0.20 days$^{-1}$. Violet plots represent the 6-body integrations and dark gray blue plots the 3-body ones. Vertical dotted lines mark the expected positions of the main mutual gravitational interactions.}
\label{fig:4}      
\end{figure*}

\begin{figure*}
\begin{tabular}{p{0.49\textwidth} p{0.5\textwidth}}
  a\includegraphics[width=0.49\textwidth]{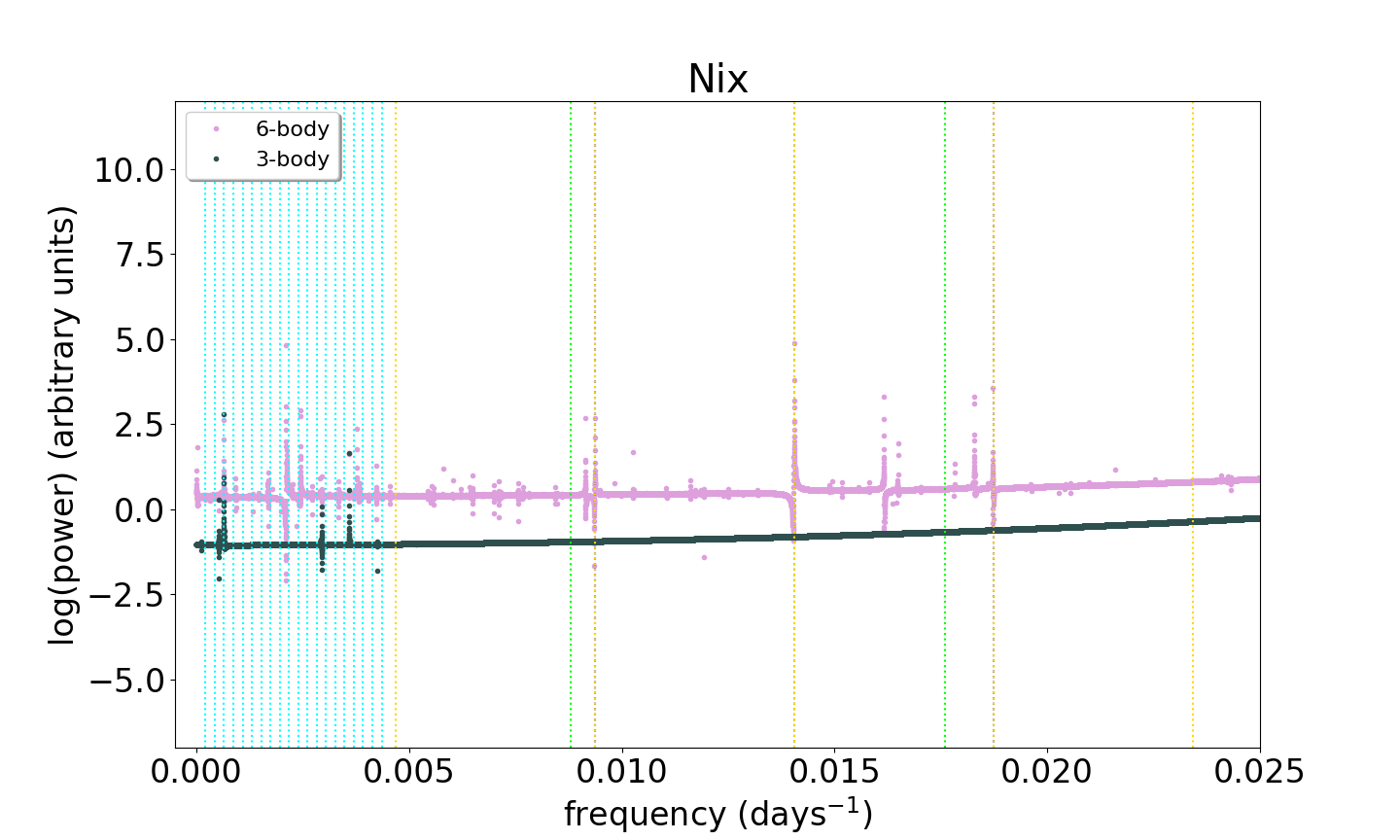}&
    b\includegraphics[width=0.49\textwidth]{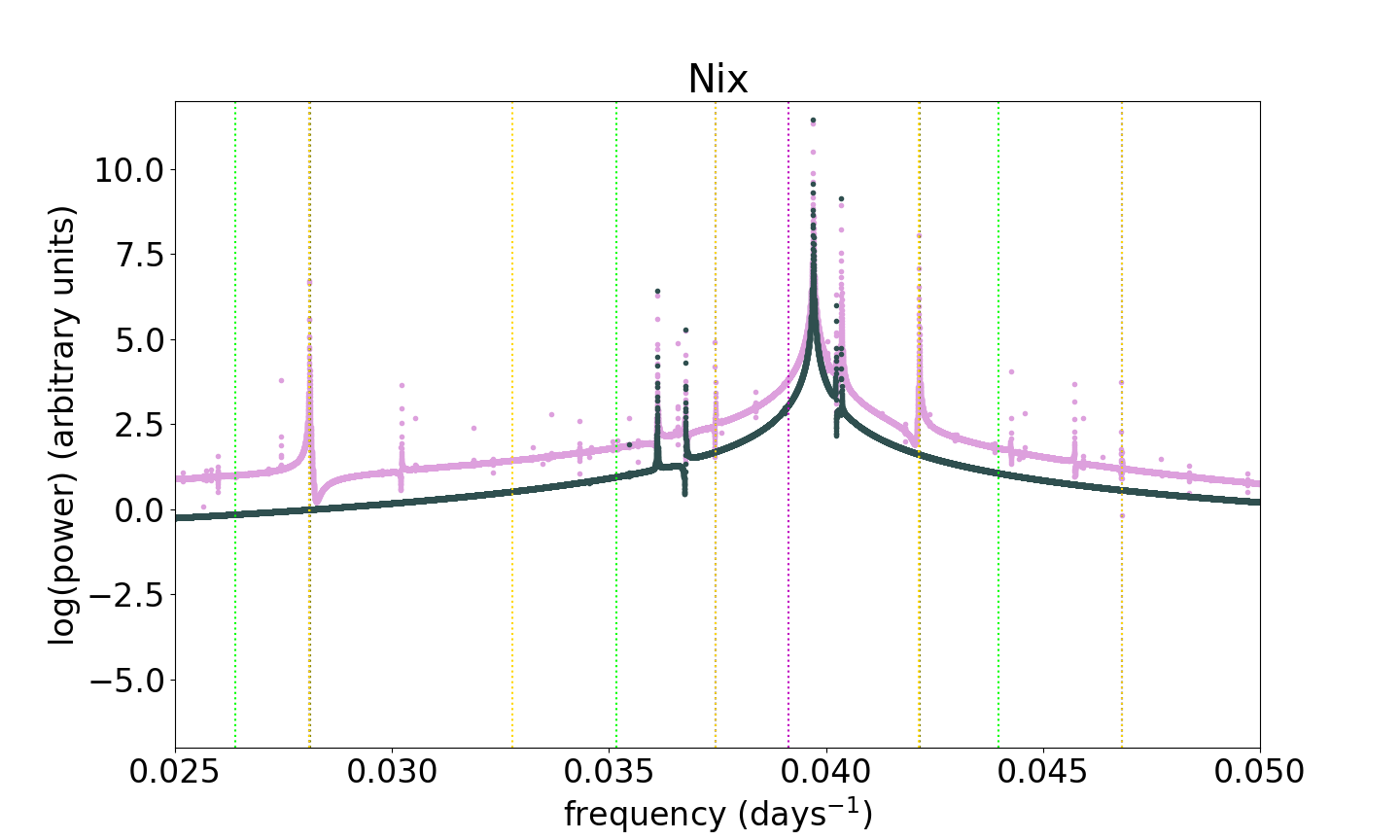} \\
     c\includegraphics[width=0.49\textwidth]{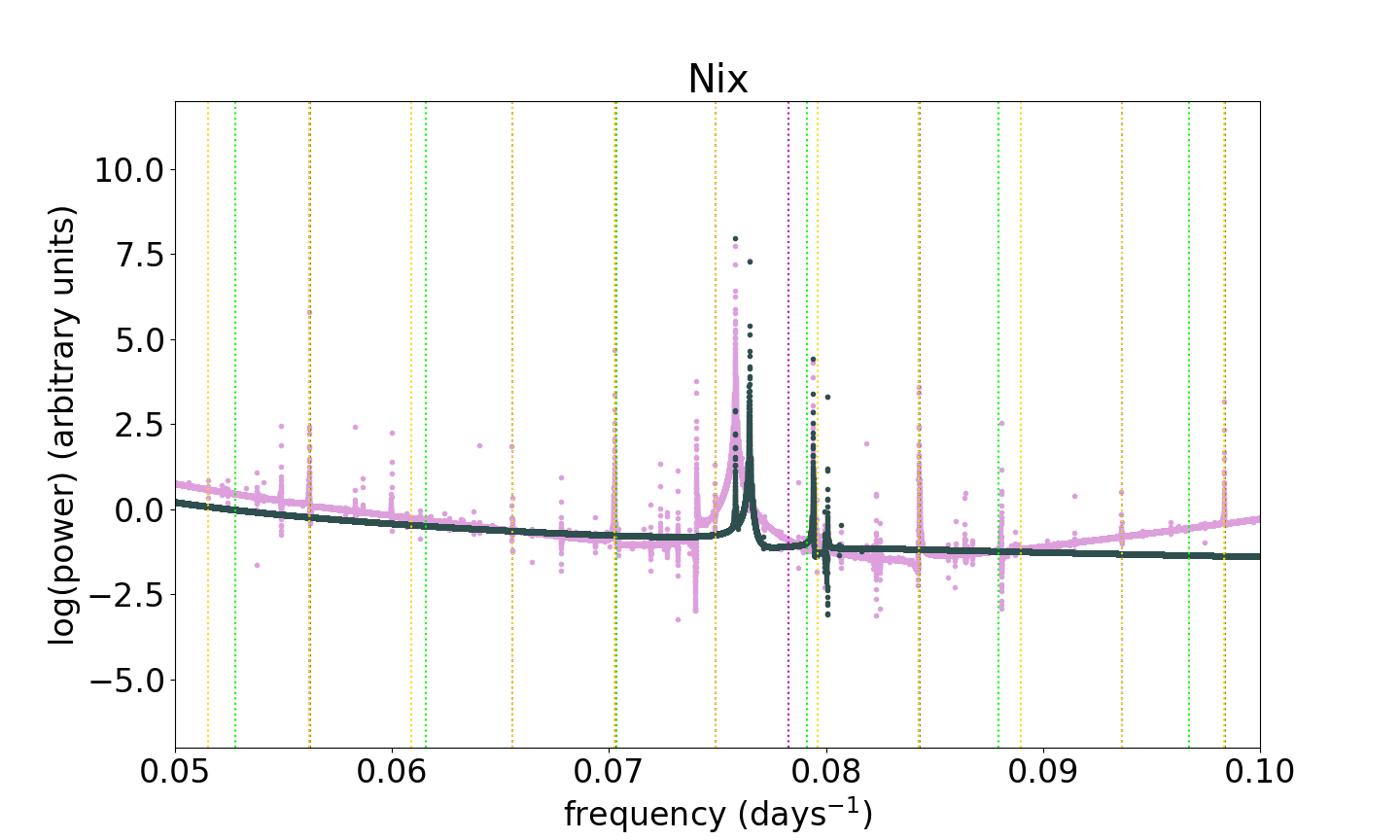} &
       d\includegraphics[width=0.49\textwidth]{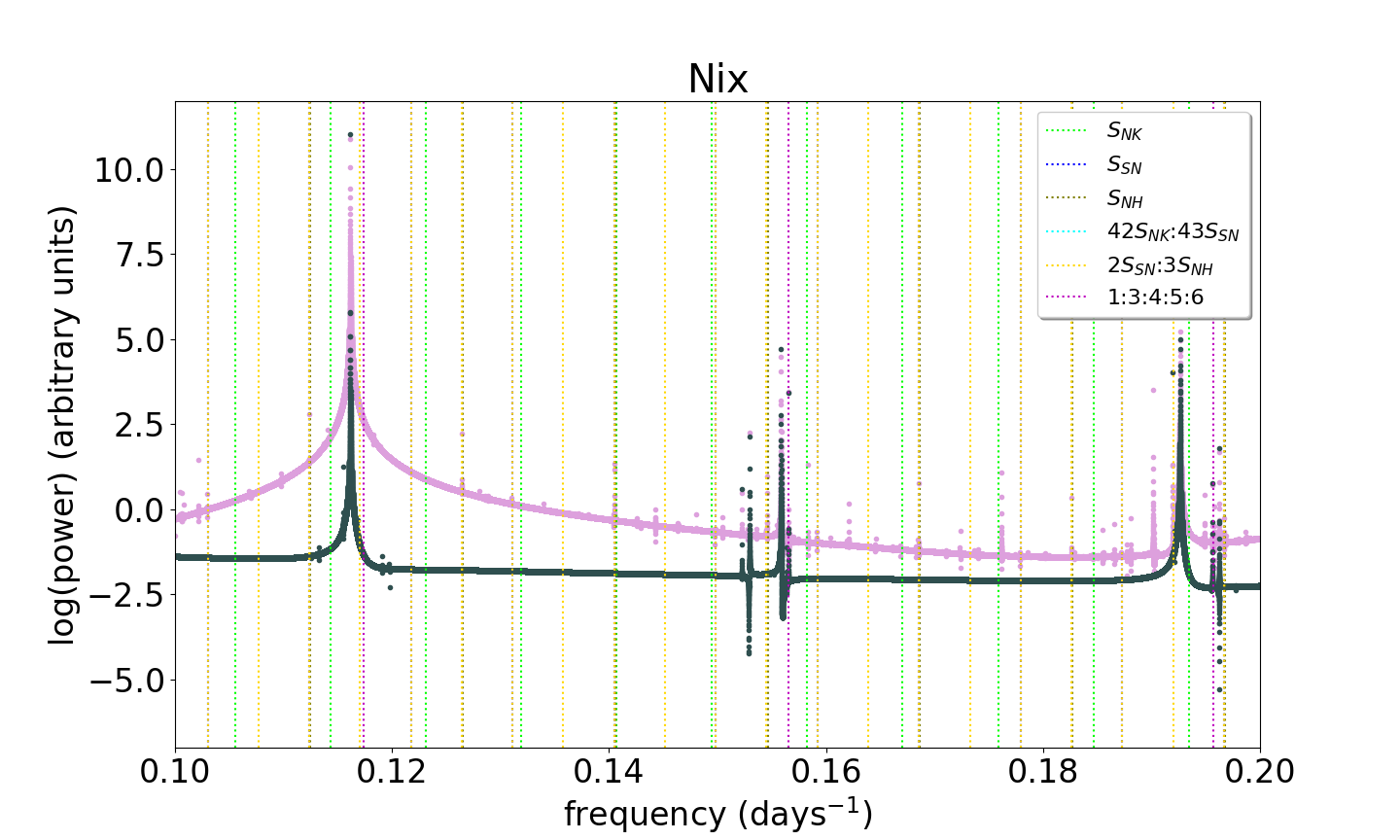} \\
     \end{tabular}
\caption{FFT successive power spectra for Nix, magnified in low frequencies. Panel a includes frequencies $\leq$ 0.0025 days$^{-1}$, panel b between 0.0025 days$^{-1}$ and 0.050 days$^{-1}$, panel c between 0.050 days$^{-1}$ and 0.10 days$^{-1}$ and panel d $\geq$ 0.20 days$^{-1}$. Violet plots represent the 6-body integrations and dark gray blue plots the 3-body ones. Vertical dotted lines mark the expected positions of the main mutual gravitational interactions.}
\label{fig:5}      
\end{figure*}

\begin{figure*}
\begin{tabular}{p{0.49\textwidth} p{0.5\textwidth}}
  a\includegraphics[width=0.49\textwidth]{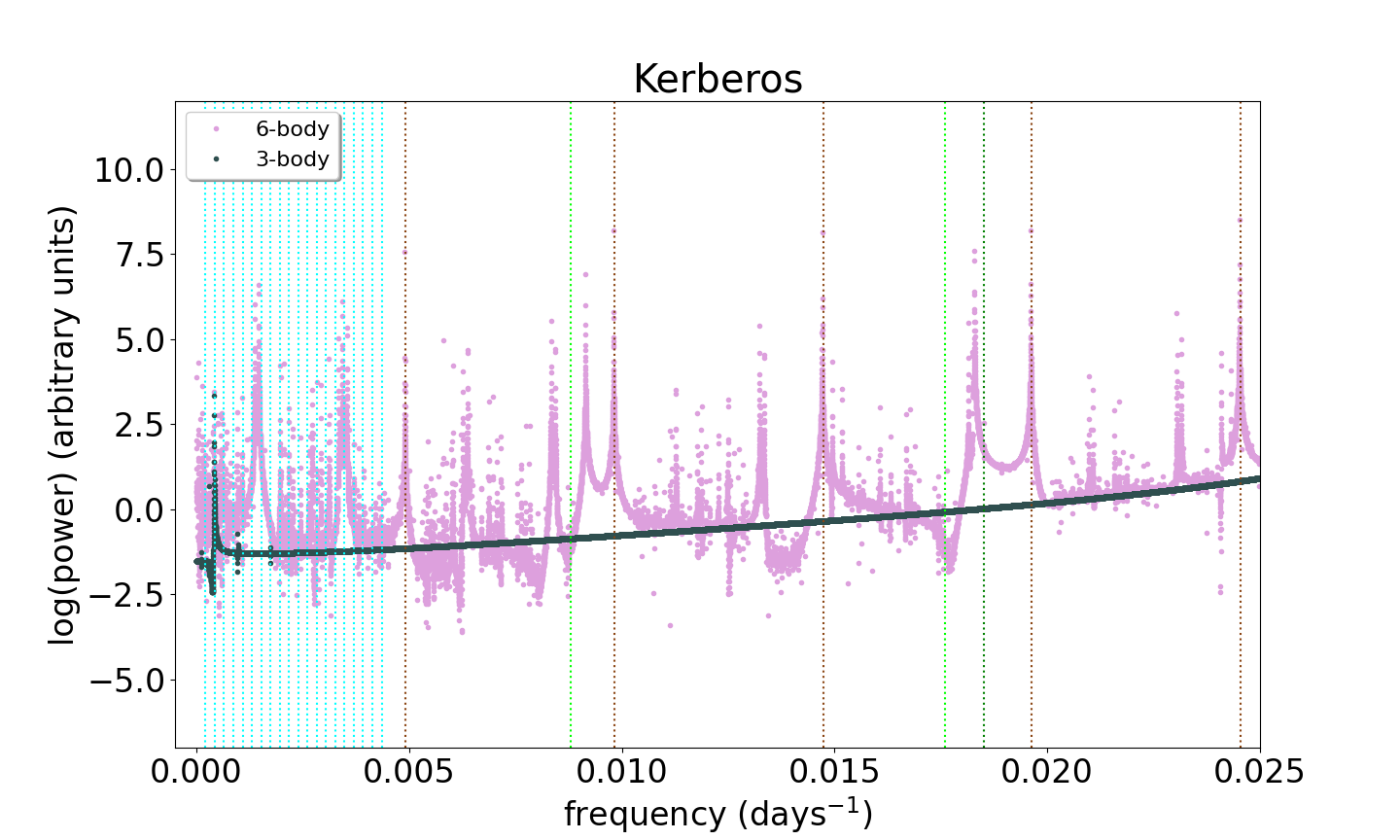}&
    b\includegraphics[width=0.49\textwidth]{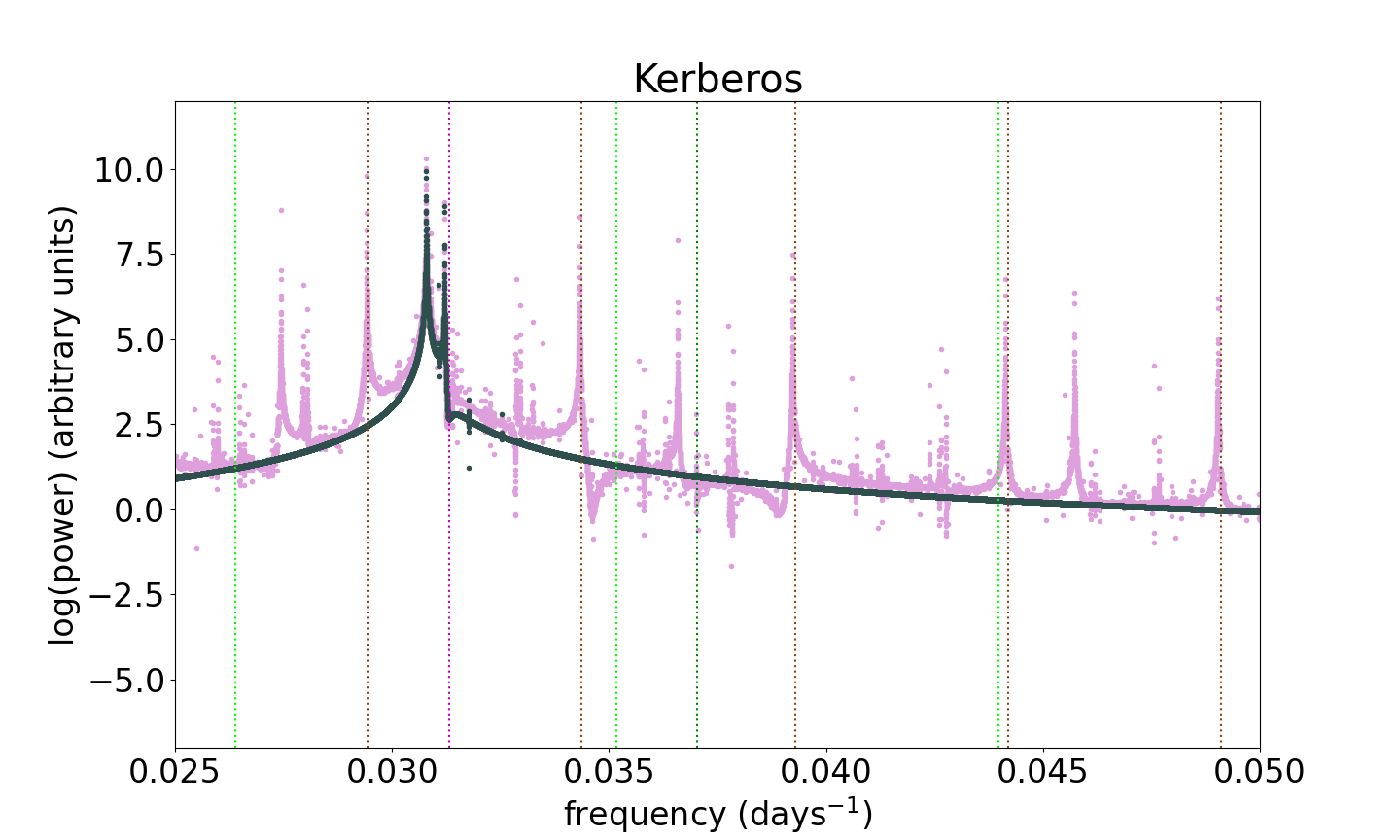} \\
     c\includegraphics[width=0.49\textwidth]{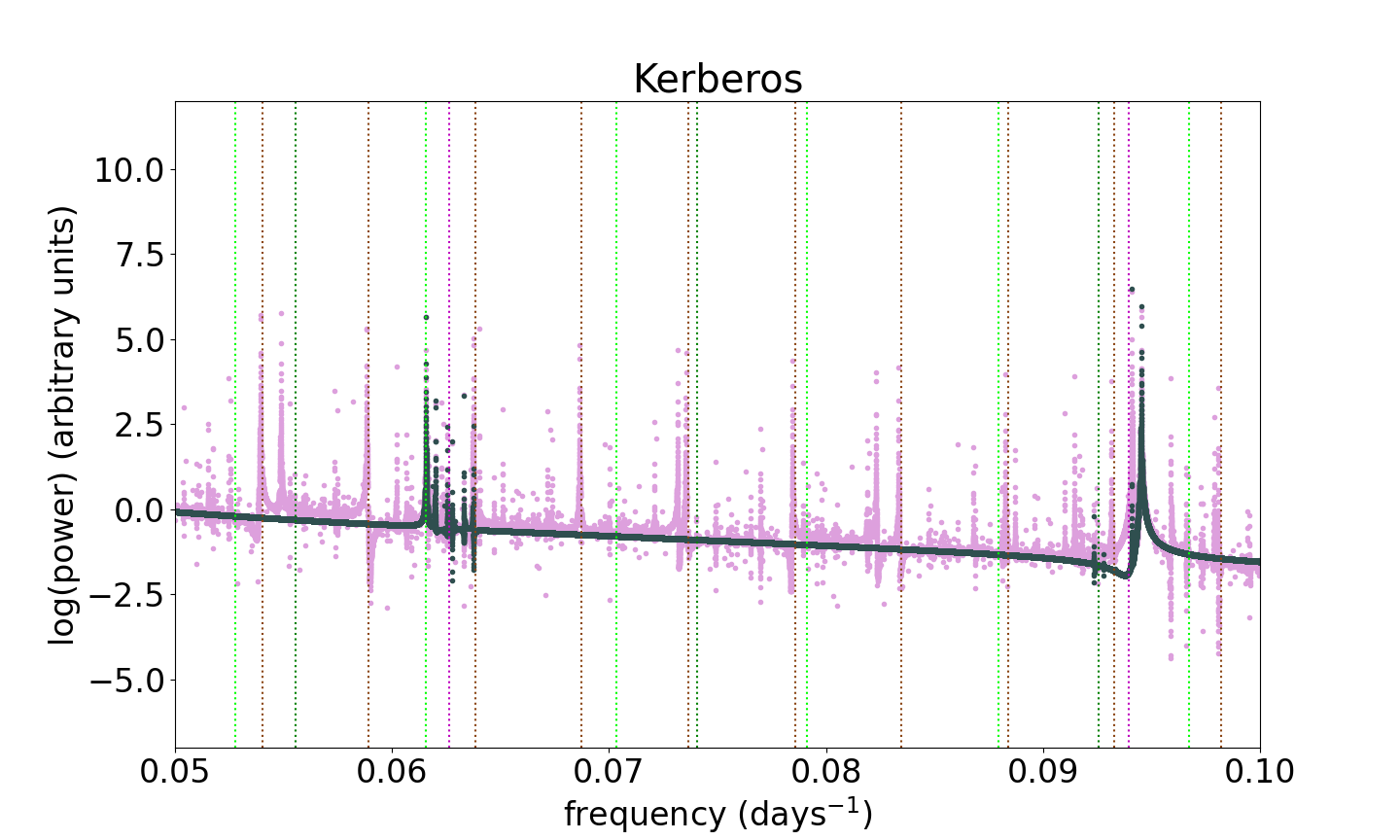} &
       d\includegraphics[width=0.49\textwidth]{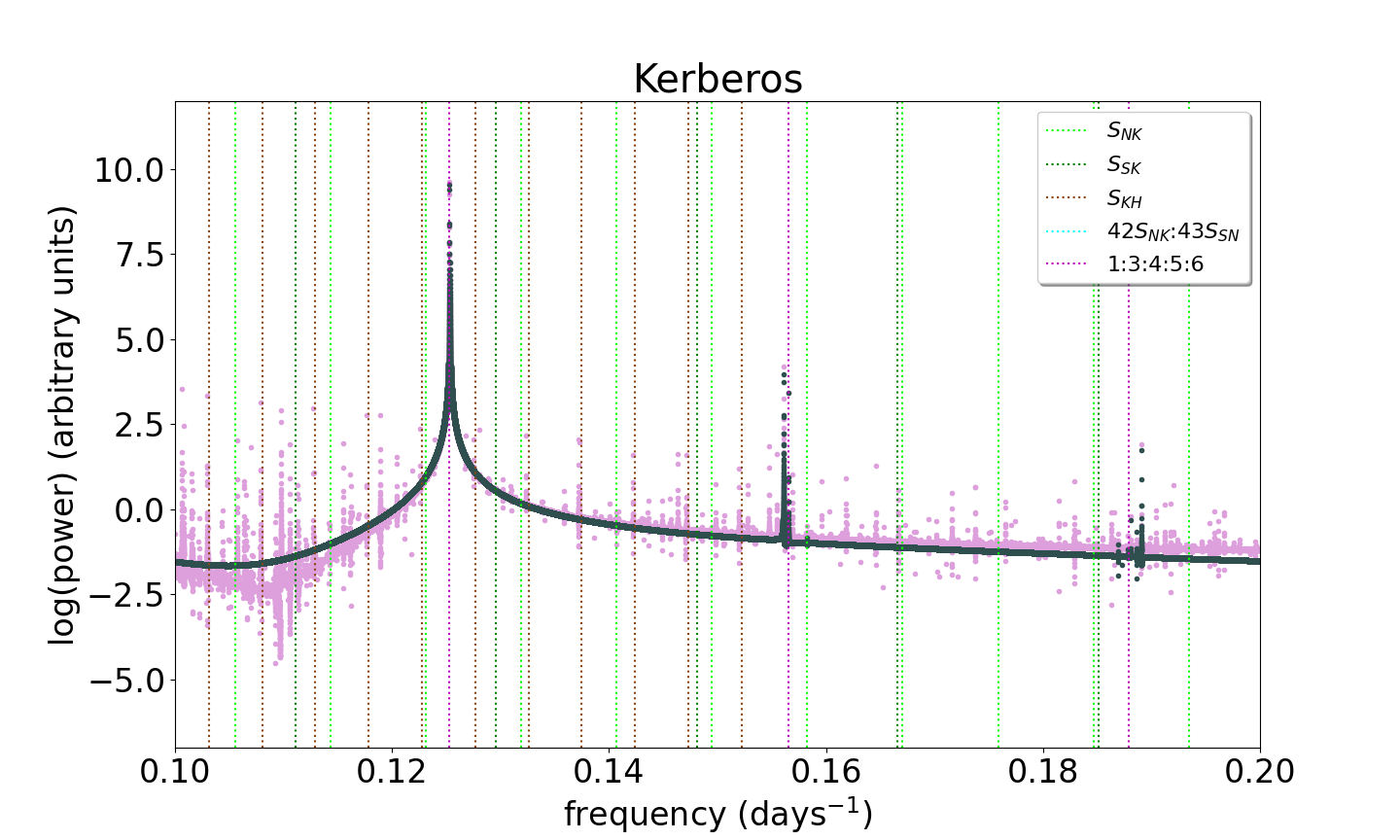} \\
     \end{tabular}
\caption{FFT successive power spectra for Kerberos, magnified in low frequencies. Panel a includes frequencies $\leq$ 0.0025 days$^{-1}$, panel b between 0.0025 days$^{-1}$ and 0.050 days$^{-1}$, panel c between 0.050 days$^{-1}$ and 0.10 days$^{-1}$ and panel d $\geq$ 0.20 days$^{-1}$. Violet plots represent the 6-body integrations and dark gray blue plots the 3-body ones. Vertical dotted lines mark the expected positions of the main mutual gravitational interactions.}
\label{fig:6}      
\end{figure*}

\begin{figure*}
\begin{tabular}{p{0.49\textwidth} p{0.5\textwidth}}
  a\includegraphics[width=0.49\textwidth]{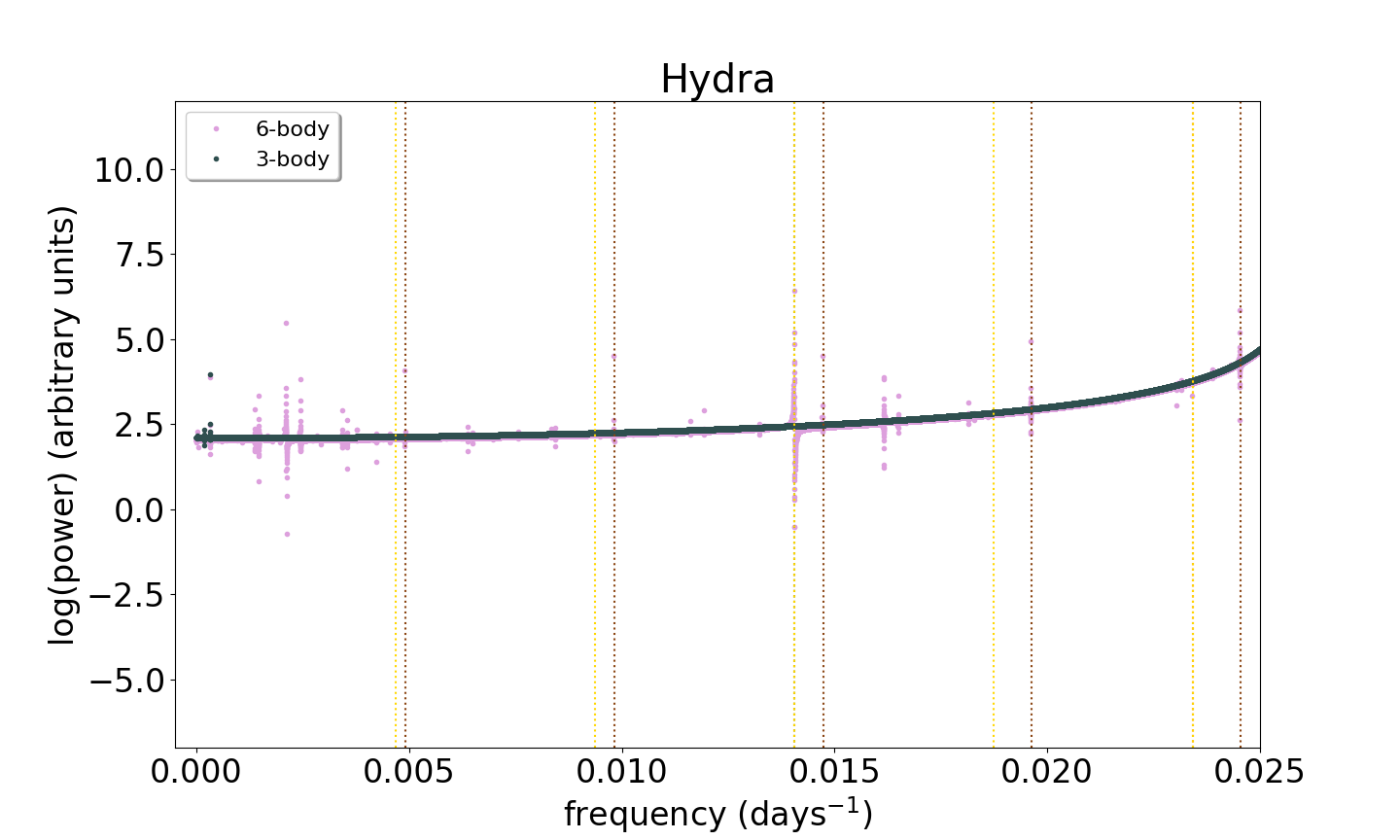}&
    b\includegraphics[width=0.49\textwidth]{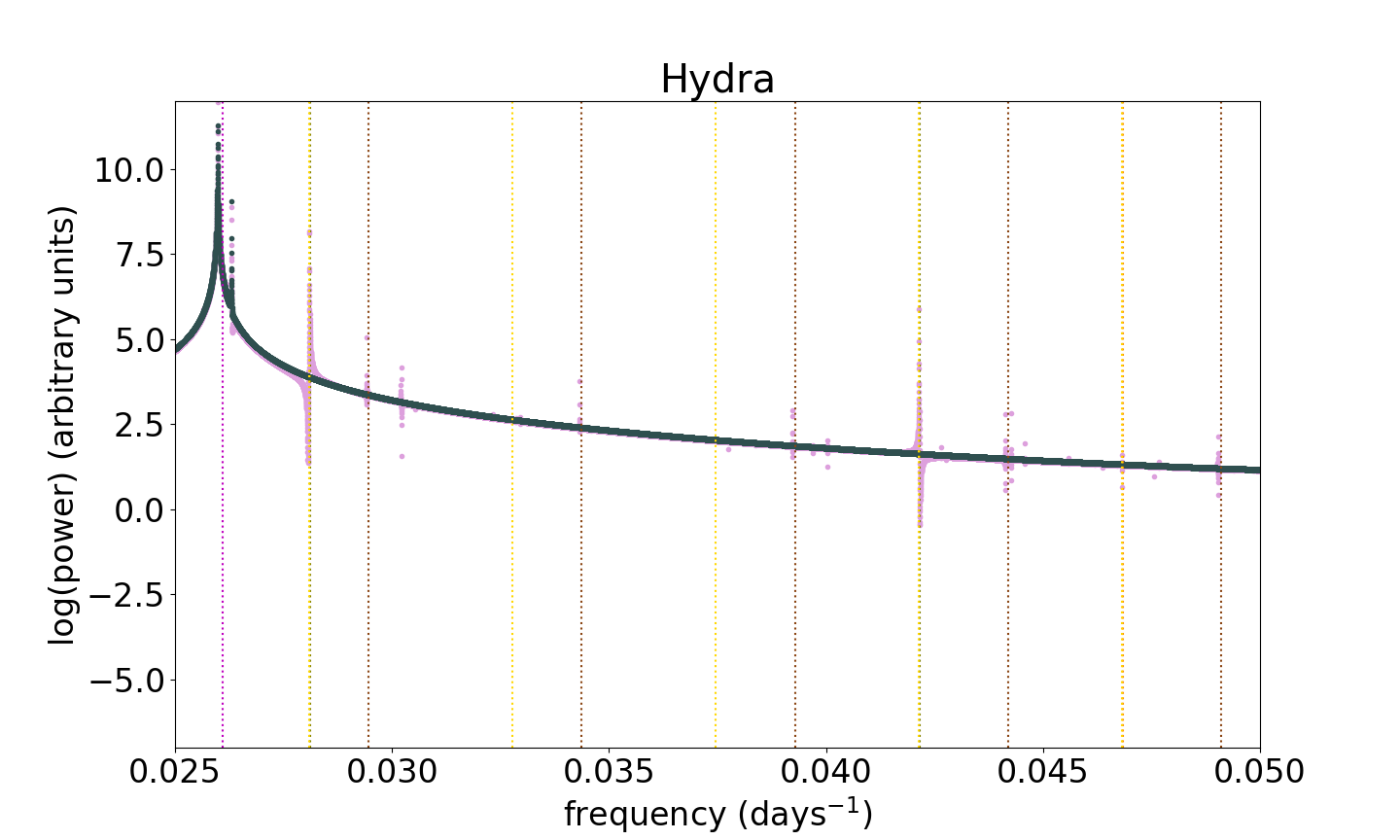} \\
     c\includegraphics[width=0.49\textwidth]{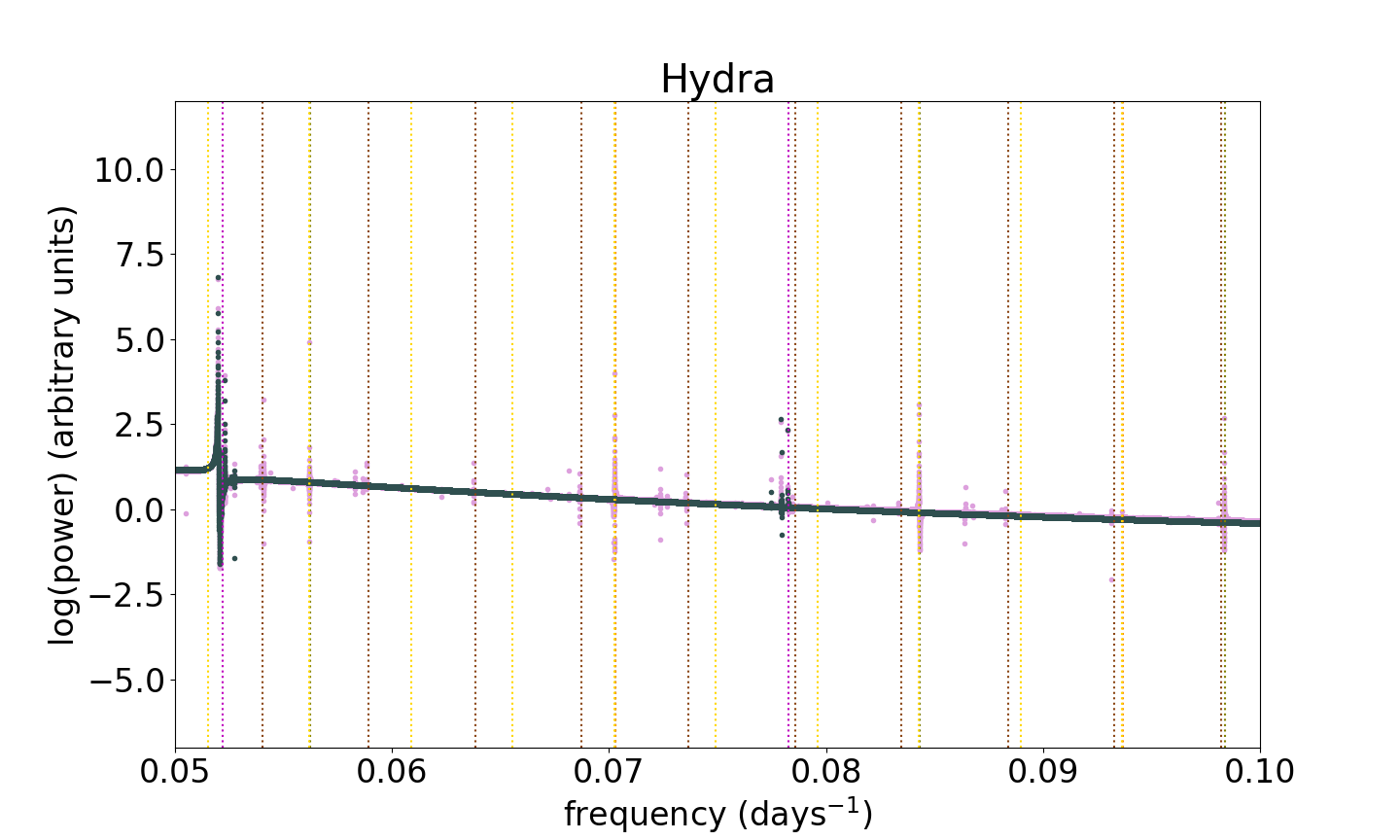} &
       d\includegraphics[width=0.49\textwidth]{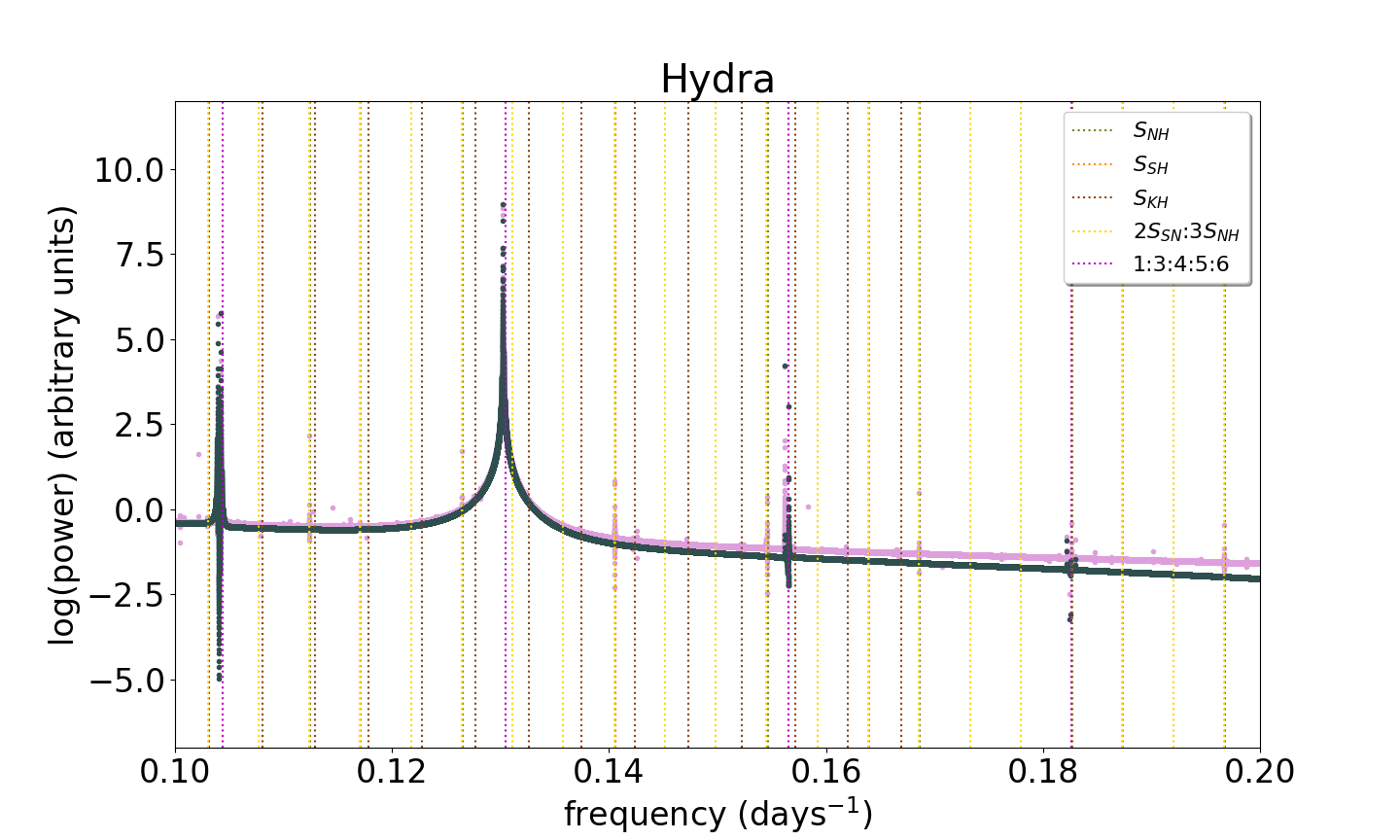} \\
     \end{tabular}
\caption{FFT successive power spectra for Hydra, magnified in low frequencies. Panel a includes frequencies $\leq$ 0.0025 days$^{-1}$, panel b between 0.0025 days$^{-1}$ and 0.050 days$^{-1}$, panel c between 0.050 days$^{-1}$ and 0.10 days$^{-1}$ and panel d $\geq$ 0.20 days$^{-1}$. Violet plots represent the 6-body integrations and dark gray blue plots the 3-body ones. Vertical dotted lines mark the expected positions of the main mutual gravitational interactions.}
\label{fig:7}      
\end{figure*}

The remaining peaks in Fig.~\ref{fig:2} at low frequencies are caused by the mutual interactions, corresponding to resonances between the moons. Considering that there are 6 bodies constituting the system, it is understandable to expect that several synodic periods (and their harmonics) can be found. In this section, we determine and identify these mutual frequencies caused by one moon to another. Some frequencies of the perturbations, for the case of Kerberos, were identified by \cite{Showalter:2015} (Extended Data Figure 3 therein). In order to separate the effects by the binary system from those by the other moons, the authors chose to merge Pluto and Charon into a single central body and compare the harmonics of resonances with the peaks of their power spectrum. In this work, we identify the mutual gravitational effects between moons by collating the simulation of the system, accounting for all objects, with a set of runs of a fictitious system where we consider the motion of each moon, accounting only for the gravitational attraction by Pluto and Charon.

Especially in frequencies near zero (i.e., long periods), an immense number of secondary peaks are visible, implying that the orbits also include long-period frequencies. Actually, in this region, the spectrum is occupied by a forest of very long frequencies. As noted above, this behavior is evident in the 6-body, long-timescale simulations. The situation is more prominent for the two lightest bodies, Styx and Kerberos. Their low masses explain their comparatively broader susceptibility to perturbations caused by the other bodies.


To quantify this effect, we provide a comparison of the power spectrum coming from the 6-body integrations and the one deriving from 3-body integrations (when simulating only the binary and one of the moons). This is shown in Fig.~\ref{fig:3}. Discrepancies are indeed more significant for Styx and Kerberos. Additionally, Fig.~\ref{fig:3} confidently establishes that the large number of peaks that appear in Fig.~\ref{fig:2} do not indicate numerical noise, but appear because of the mutual gravitational effects. That is, the long-term periodicity terms of one moon to each other cover almost entirely the frequency region near zero. In 3-body simulations, especially this area of the spectrum lacks peaks 
because only moon-binary interactions are considered.

In order to study more precisely the mutual gravitational interactions, we zoom in the area of low frequencies (<0.20 days$^{-1}$) of the frequency spectra in Fig.~\ref{fig:3}. These magnified spectra are presented successively in Figs.~\ref{fig:4}--\ref{fig:7} for Styx, Nix, Kerberos and Hydra, respectively. For a more efficient visualization, we divide the low-frequency area of each moon into four panels, corresponding to frequencies $\leq$ 0.0025 days$^{-1}$(panels a), between 0.0025 days$^{-1}$ and 0.050 days$^{-1}$ (panels b), between 0.050 days$^{-1}$ and 0.10 days$^{-1}$(panels c), and 
between 0.10 days$^{-1}$ and
0.20 days$^{-1}$ (panels d). That way we managed to study in detail the reciprocal effects that gradually become weaker compared to the forced oscillations as the frequency rises. Obviously, when improving the resolution of the periodogram, separate oscillations arise from the seemingly almost continuous spectrum of frequencies in Fig.~\ref{fig:3}. Some of these frequencies, though immense in number, drop in favor of a few more dominant peaks.

At first, all four small moons are placed close to mean motion resonances (MMRs) with Charon. Specifically, the ratios of their Keplerian orbital periods are about 1:3:4:5:6, similar to the Laplace configuration in the Galilean moons of Jupiter. Although they do not definitely belong in the resonance according to the currently accepted orbital elements for the four moons \citep{Brozovic:2015,Showalter:2015}, adopting the range of their uncertainties could certainly place them well inside the MMRs \citep{Giuppone:2022}. Additionally, a resonant term could affect the tidal damping of a moon even if it is not located in the actual position of the resonance, as shown, for example, for the 3:1 resonance of Nix in \cite{Lithwick:2008}. For that reason, we examine this type of resonance in our analysis. In the same study it is shown that even other MMRs of the form N:1 with Charon could be significant for some moons (e.g., the 2:1 MMR of Nix with Charon). Nevertheless, in order to maintain our analysis feasible and focused on the most strongest mutual interactions, we examine only the most prominent 1:3:4:5:6 resonance with Charon. For each moon, the anticipated positions of the harmonics of this resonances is shown in purple vertical dotted lines in the frequency spectra of Figs.~\ref{fig:4}--\ref{fig:7}.

More complex resonances can also be defined. After an extensive search for potential resonances, \cite{Showalter:2015} found two such major resonances implicating three moons. The strongest resonance identified was $\Phi = 3 \lambda_S -5 \lambda_N + 2 \lambda_H \approx 180^{\circ}$, which implies that the synodic period of Nix-Hydra divided by that of Styx-Nix is 3:2, i.e., 3S$_{NH} = $ 2S$_{SN}$, where the subscripts note the respective moons. A second resonance, now involving Styx, Nix and Hydra, was found as 42S$_{NK} \approx $ 43S$_{SN}$. We calculate the frequencies induced by the above resonances and indicate their harmonics with yellow and cyan dotted lines in the frequency spectra.

Of course, apart from any resonances we note the strong interaction of one with another over a time period equal to synodic period of them. Therefore, we calculate the respective frequencies by the synodic frequencies $S_{SN}$, $S_{SK}$, $S_{SH}$, $S_{NK}$, $S_{NH}$, and $S_{KH}$. The positions of their harmonics are shown in blue, green, orange, lime, olive and brown vertical dotted lines in Fig.~\ref{fig:4}--\ref{fig:7}, respectively.

To avoid further contamination of the images with many vertical lines representing expected oscillatory modes, in these figures we present only the frequencies by the mutual interactions (the forced frequencies by the binary system are shown clearly in Figs.~\ref{fig:1} and~\ref{fig:2}). Peaks where the 6-body spectra match with the 3-body spectra mark the modes by Pluto and Charon, as explained in Section \ref{sec:3.1}. As far as the long-period resonance 42S$_{NK} \approx $ 43S$_{SN}$ is considered, we show only its first 20 harmonics because additional lines could heavily fill the figures. 

Even though each mode would ideally appear as delta function (negligible width), in practise many of them have a significant width. As is evident from the vertical lines in Figs.~\ref{fig:4}--\ref{fig:7}, this is not primarily an effect of the limited accuracy of the FFT method. Instead, it is a result of a number of oscillations with very close frequencies.

As noted by \cite{Gakis:2022}, Keplerian osculating elements are not sufficient to describe circumbinary orbits. 
Hence, obtaining the synodic periods of one moon with another or the locations of the resonances is uncertain, as it is based on their deduced orbital periods. Future observations may increase the accuracy of the measurements providing more robust estimates for the masses and the 3D position and velocity vectors of the objects. This is yet another reason for possible minor discrepancies between a peak in the periodogram and the anticipated position of its respective frequency (no more than $\sim$0.2\%). Apart from that, these orbital elements themselves have not been decisively determined yet, as we find significant differences in previous data tables. For our calculations, we adopted the values in Table~\ref{tab:1}, deduced by \cite{Showalter:2015}.

Styx is the moon most heavily affected by mutual perturbations, along with Kerberos. The strongest mode in Styx's spectrum (Fig.~\ref{fig:4}) is the one induced by Nix, which is the closest moon. The interaction with Hydra has a similar strength, since Hydra is the most massive moon (though most distant to Styx). It is then evident that the resonance 3S$_{NH} = $ 2S$_{SN}$ produces the most important perturbations in the orbital pattern of Styx. On the other hand, due to its low mass, Kerberos forces much weaker modes to Styx, about 3 times weaker compared to the perturbations by Nix and Hydra. The purple lines in Fig.~\ref{fig:4}, showing the resonance 1:3:4:5:6, correspond to small peaks either, as Styx does not belong well within the resonance. On the contrary, Styx experiences the effects of the resonance 42S$_{NK} \approx $ 43S$_{SN}$ strongly at low frequencies, but as the frequency of their harmonics rises, their strength gradually drops. This is, however, the reason for the large number of peaks in between the most prominent ones, by 3S$_{NH} = $ 2S$_{SN}$, though they are not shown in detail in the diagrams.

Nix appears to have a similar behavior, though the number and strength of peaks is not as large as Styx's. Again, the resonance 3S$_{NH} = $ 2S$_{SN}$ is the dominant one, while there is a slight preference to the modes by Hydra rather than those by Styx. Despite their proximity, the gravitational interaction between Nix and Kerberos is not powerful; Hydra has a much clearer imprint in the frequency spectrum of Nix than Kerberos. The resonance $42S_{NK} \approx $ 43S$_{SN}$ mainly affects the low-frequency region, as noted by the cyan vertical lines in Fig.~\ref{fig:5}, whereas the 4:1 MMR with Charon does not seem to produce an intense peak.

The strongest peaks in the spectrum of Kerberos (see Fig.~\ref{fig:6}) are undeniably the harmonics of the synodic frequency $S_{KH}$. They are followed by the interactions of Kerberos with Styx, while the mutual effects between Kerberos and Nix seem to be minuscule compared to the above two. However, the 5:1 MMR with Charon has a more substantial impact on Kerberos than the effect N:1 MMR has on Styx or Nix. The resonance 42S$_{NK} \approx $ 43S$_{SN}$ remains quite strong even for higher frequencies as well, forming evident peaks in between the main components of the synodic frequencies.

Hydra is the most massive moon and has the largest distance from the system's barycenter. Evidently the effects by either Pluto--Charon or the other three moons are the lowest, and its motion is the closest approximation to a typical Keplerian elliptic orbit in the system. As Fig.~\ref{fig:7} suggests, Hydra is mainly disturbed by the resonance 3S$_{NH} = $ 2S$_{SN}$. The motion of Kerberos also induces some perturbations, though of a smaller scale. Apart from this, Hydra's resonance with Charon is evident through some intermediate peaks.

For each one of the moons there are a number of lower-amplitude peaks that are not explained by the resonances examined above. This effect is more obvious for the lightest moons, Styx and Kerberos. We argue that their origin lies in the N:1 MMRs with Charon. As mentioned, a specific type of resonance could have a gravitational effect on a moon even if it is not located in the exact position of the resonance. Consequently, our conclusion is that these smaller-scale modes are created by near-resonance kicks other than 1:3:4:5:6.

Lastly, a significant remark on the power spectra of the four moons is that the peaks by mutual interactions have a comparable size to the binary-induced peaks for Styx and Kerberos, at least in the low-frequency region. This is not the case for the heavier Nix and Hydra, where the binary effects are in general stronger in the entire frequency region. This implies that the reciprocal effects for the less massive moons may influence the orbits just as much as the binary system, or even dominate over them for long periods. Consequently, we speculate that modeling the orbits of these moons would not be realistic when ignoring the dynamics of the rest of the objects in the system.

\section{Conclusions}
\label{sec:4} 

In this paper we identified the most prominent frequencies at which the moons of Pluto and Charon swing. To achieve this, we employed FFTs to computed orbital elements by a semi-analytic and an arithmetic process. It is confirmed that forced oscillations caused by the rotating non-axisymmetric components of the central binary extend to values set by the $C_k$ term and occur at frequencies $k(n_S-n_{PC})$. We also notice that although our adopted linearized theory allows it, minimizing such frequencies is impossible even when demanding $e_{free}=0$ for the circumbinary orbits, because of the ubiquitous nonlinear terms of gravity. 

By collating outcomes from restricted 3-body simulations along with the case in which every moon is present, we also managed to assess the mutual gravitational effect. We deduce that the low-mass moons Styx and Kerberos are more intensely affected by the other objects, and over long integration times the mutual interactions have a significant effect on their orbital frequencies. In fact, the mutual effects concerning these two moons have comparable (in strength) perturbations to the orbits with the central binary, at least in the low-frequency region.

Specifically, we found that the strongest mutual gravitational interactions are caused by the resonances 3S$_{NH} = $ 2S$_{SN}$and 42S$_{NK} \approx $ 43S$_{SN}$. In the first case, strong kicks are sparsely produced, while the second, longer-period, type of resonance fills the region in between. Nonetheless, the N:1 MMR with Charon is not as powerful, which is attributed to the proximity of the moons to the resonance.

\begin{acknowledgements}
      The authors thank Prof. Alain Vienne for useful suggestions that enhanced this work. The numerical code used in this work was branched from an n-body code (https://github.com/pmocz/nbody-python) created by Philip Mocz. KNG acknowledges support by grant University of Patras, ELKE81641.
\end{acknowledgements}

%
%

\bibliographystyle{aa} 
\bibliography{bibtex.bib} 

\providecommand{\noopsort}[1]{}\providecommand{\singleletter}[1]{#1}%
\begin{thebibliography}{19}
\expandafter\ifx\csname natexlab\endcsname\relax\def\natexlab#1{#1}\fi

\bibitem[{{Bromley} \& {Kenyon}(2021)}]{Bromley-Kenyon:2020}
{Bromley}, B.~C. \& {Kenyon}, S.~J. 2021, \aj, 161, 25

\bibitem[{{Brozovi{\'c}} {et~al.}(2015){Brozovi{\'c}}, {Showalter}, {Jacobson},
  \& {Buie}}]{Brozovic:2015}
{Brozovi{\'c}}, M., {Showalter}, M.~R., {Jacobson}, R.~A., \& {Buie}, M.~W.
  2015, \icarus, 246, 317

\bibitem[{{Buie} {et~al.}(2012){Buie}, {Tholen}, \& {Grundy}}]{Buie:2012}
{Buie}, M.~W., {Tholen}, D.~J., \& {Grundy}, W.~M. 2012, \aj, 144, 15

\bibitem[{{Gakis} \& {Gourgouliatos}(2022)}]{Gakis:2022}
{Gakis}, D. \& {Gourgouliatos}, K.~N. 2022, Celestial Mechanics and Dynamical
  Astronomy, 134, 14

\bibitem[{{Gakis} \& {Gourgouliatos}(2023)}]{Gakis:2022bb}
{Gakis}, D. \& {Gourgouliatos}, K.~N. 2023, \mnras, 519, 3832

\bibitem[{{Georgakarakos} \& {Eggl}(2015)}]{Georgakarakos:2015}
{Georgakarakos}, N. \& {Eggl}, S. 2015, \apj, 802, 94

\bibitem[{Giuppone {et~al.}(2022)Giuppone, Rodr{\'\i}guez, Michtchenko, \&
  de~Almeida}]{Giuppone:2022}
Giuppone, C.~A., Rodr{\'\i}guez, A., Michtchenko, T.~A., \& de~Almeida, A.~A.
  2022, Astronomy \& Astrophysics, 658, A99

\bibitem[{Kenyon \& Bromley(2019{\natexlab{a}})}]{Kenyon:2019b}
Kenyon, S.~J. \& Bromley, B.~C. 2019{\natexlab{a}}, The Astronomical Journal,
  158, 69

\bibitem[{Kenyon \& Bromley(2019{\natexlab{b}})}]{Kenyon:2019a}
Kenyon, S.~J. \& Bromley, B.~C. 2019{\natexlab{b}}, The Astronomical Journal,
  157, 79

\bibitem[{{Kenyon} \& {Bromley}(2022)}]{2022AJ....163..238K}
{Kenyon}, S.~J. \& {Bromley}, B.~C. 2022, \aj, 163, 238

\bibitem[{Laskar(1999)}]{Laskar:1999}
Laskar, J. 1999, in Hamiltonian systems with three or more degrees of freedom
  (Springer), 134--150

\bibitem[{{Lee} \& {Peale}(2006)}]{Lee:2006}
{Lee}, M.~H. \& {Peale}, S.~J. 2006, \icarus, 184, 573

\bibitem[{{Leung} \& {Lee}(2013)}]{Leung:2013}
{Leung}, G. C.~K. \& {Lee}, M.~H. 2013, \apj, 763, 107

\bibitem[{Lithwick \& Wu(2008)}]{Lithwick:2008}
Lithwick, Y. \& Wu, Y. 2008, arXiv preprint arXiv:0802.2939

\bibitem[{{Michaely} {et~al.}(2017){Michaely}, {Perets}, \&
  {Grishin}}]{Michaely:2017}
{Michaely}, E., {Perets}, H.~B., \& {Grishin}, E. 2017, \apj, 836, 27

\bibitem[{{Showalter} \& {Hamilton}(2015)}]{Showalter:2015}
{Showalter}, M.~R. \& {Hamilton}, D.~P. 2015, \nat, 522, 45

\bibitem[{{Stern} {et~al.}(2015){Stern}, {Bagenal}, {Ennico}, {Gladstone},
  {Grundy}, {McKinnon}, {Moore}, {Olkin}, {Spencer}, {Weaver}, {Young},
  {Andert}, {Andrews}, {Banks}, {Bauer}, {Bauman}, {Barnouin}, {Bedini},
  {Beisser}, {Beyer}, {Bhaskaran}, {Binzel}, {Birath}, {Bird}, {Bogan},
  {Bowman}, {Bray}, {Brozovic}, {Bryan}, {Buckley}, {Buie}, {Buratti},
  {Bushman}, {Calloway}, {Carcich}, {Cheng}, {Conard}, {Conrad}, {Cook},
  {Cruikshank}, {Custodio}, {Dalle Ore}, {Deboy}, {Dischner}, {Dumont},
  {Earle}, {Elliott}, {Ercol}, {Ernst}, {Finley}, {Flanigan}, {Fountain},
  {Freeze}, {Greathouse}, {Green}, {Guo}, {Hahn}, {Hamilton}, {Hamilton},
  {Hanley}, {Harch}, {Hart}, {Hersman}, {Hill}, {Hill}, {Hinson}, {Holdridge},
  {Horanyi}, {Howard}, {Howett}, {Jackman}, {Jacobson}, {Jennings}, {Kammer},
  {Kang}, {Kaufmann}, {Kollmann}, {Krimigis}, {Kusnierkiewicz}, {Lauer}, {Lee},
  {Lindstrom}, {Linscott}, {Lisse}, {Lunsford}, {Mallder}, {Martin}, {McComas},
  {McNutt}, {Mehoke}, {Mehoke}, {Melin}, {Mutchler}, {Nelson}, {Nimmo},
  {Nunez}, {Ocampo}, {Owen}, {Paetzold}, {Page}, {Parker}, {Parker},
  {Pelletier}, {Peterson}, {Pinkine}, {Piquette}, {Porter}, {Protopapa},
  {Redfern}, {Reitsema}, {Reuter}, {Roberts}, {Robbins}, {Rogers}, {Rose},
  {Runyon}, {Retherford}, {Ryschkewitsch}, {Schenk}, {Schindhelm}, {Sepan},
  {Showalter}, {Singer}, {Soluri}, {Stanbridge}, {Steffl}, {Strobel}, {Stryk},
  {Summers}, {Szalay}, {Tapley}, {Taylor}, {Taylor}, {Throop}, {Tsang},
  {Tyler}, {Umurhan}, {Verbiscer}, {Versteeg}, {Vincent}, {Webbert}, {Weidner},
  {Weigle}, {White}, {Whittenburg}, {Williams}, {Williams}, {Williams},
  {Woods}, {Zangari}, \& {Zirnstein}}]{Stern:2015}
{Stern}, S.~A., {Bagenal}, F., {Ennico}, K., {et~al.} 2015, Science, 350,
  aad1815

\bibitem[{Sutherland \& Kratter(2019)}]{Sutherland:2019}
Sutherland, A.~P. \& Kratter, K.~M. 2019, Monthly Notices of the Royal
  Astronomical Society, 487, 3288

\bibitem[{{Woo} \& {Lee}(2020)}]{Woo:2020}
{Woo}, J. M.~Y. \& {Lee}, M.~H. 2020, \aj, 159, 277

\end{thebibliography}

\end{document}